\documentclass[12pt]{article}

\usepackage{graphicx}

\begin{document}

\hfill{forarxiv/He4.tex}

\hrule

\vspace{.3cm}

\centerline {\Large \bf Macro-orbitals and Microscopic Theory of }

\smallskip
\centerline {\Large \bf a System of Interacting Bosons}

\vspace{0.5cm}

\centerline {\bf Yatendra S. Jain}

\vspace{.3cm} 

\centerline {Department of Physics, North-Eastern Hill University,} 

\centerline {Shillong-793 022, Meghalaya, India}

\begin{abstract}       
Macro-orbital representation of a particle in a many body
system is used to develop the microscopic theory of a
system of interacting bosons.  Each particle in the system
represents a pair of particles moving with equal and
opposite momenta $({\bf q}, -{\bf q})$ at their center of
mass (CM) which moves with momentum {\bf K} in the
laboratory frame.  Below certain temperature (say,
$T_{\lambda}$), these particles assume a state of
$({\bf q}, -{\bf q})$ bound pairs (named as SMW pairs).
The $\lambda-$transition is found to be a consequence of
inter-particle quantum correlations clubbed with zero-point
repulsion and inter-particle attraction; it is an onset of
the {\it order-disorder} of particles in their $\phi$-space
followed simultaneously by their {\it Bose Einstein
condensation} as SMW pairs in a state of $q = q_o = {\pi}/d$
and $K=0$.  Particles at $T \le T_{\lambda}$ acquire
collective binding which locks them at $<k> = 0$ in momentum
space, $<r> = \lambda/2$ in real space and at $\Delta\phi
= 2n\pi$ (with $n$  = 1, 2, 3, ...) in $\phi-$space.
Consequently, the entire system assumes {\it mechanical
strain} in inter-particle bonds and behaves like a single
macroscopic molecule.  The collective binding is identified
as an energy gap between the superfluid and normal fluid
phases.  The fractional density of condensed particles
($n_{K=0}(T)$) varies smoothly from $n_{K=0}(T_{\lambda})
= 0$ to $n_{K=0}(0) = 1.0$.  The $\lambda-$transition
represents the occurrence of twin phenomena of broken gauge
symmetry and phase coherence.  In variance with the
conventional belief, the system does not have single
particle $p=0$ condensate.  In addition to the well known
modes of collective motions (such as phonons), the
superfluid also exhibits a new kind of quantum
quasi-particle, {\it omon}, (a phononlike wave of the
oscillations of momentum coordinates of the particles.
Omons are sustained because the strain energy of
inter-particle bonds is a function of relative momentum
of particles.  The theory explains the properties of
$He-II$, including the origin of quantized vortices,
critical velocities, logarithmic singularity of
specific heat and related properties at quantitative level.
It conforms to the excluded volume
condition, microscopic and macroscopic uncertainty, and
renders microscopic foundation to two fluid theory of
Landau, $\Psi$-theory of Ginzburg, the idea of macroscopic
wave  function of London, {\it etc}.   The framework of
this theory also helps in unifying the physics of widely
different systems of interacting bosons and fermions.

\end{abstract}

{\it email} :  ysjain@email.com

Key-words : Macro-orbitals, Microscopic theory, Superfluidity,
Liquid helium-4, Bosons.

PACS : 67.20.+k, 67.40.-w, 67.40.Db, 67.40.Db,

\bigskip

\centerline{\bf 1.0  INTRODUCTION}

\bigskip
{\it Liquid} $^4He$ (LHE-4), - a typical {\it system of
interacting bosons} (SIB), has been investigated extensively
for its unique properties at {\it low temperature} (LT) and
results are reviewed in several articles ({\it e.g.} [1-10]
and other references cited therein). It transforms from its
normal (N) phase $(He-I)$ to superfluid (S) phase $(He-II)$
at $T_{\lambda} = 2.17$ K with the latter exhibiting several
unique properties including superfluidity.  One finds that
these properties can not be explained as the properties of a
classical liquid and the same is true for the fact that
helium remains liquid down to the lowest attainable {\it
temperature} (T).  Consequently, these properties are,
rightly, related to the manifestation of wave nature of
particles at macroscopic level and LHE-4 provides unique
opportunity to study this aspect.  Naturally, different
studies of a SIB are considered to be of fundamental
importance.

\bigskip
Soon after the discovery of superfluidity of $He-II$ [11],
London [12] proposed that the phenomenon arises from
$p=0$ {\it condensate}, -a macroscopically large fraction
($n_{p=0}(T)$) of $^4He$ atoms occupying a single particle
state of momentum $p=0$.  To this effect, London was
guided by the well known conclusion that a {\it system of
non-interacting bosons} (SNIB) should exhibit a transition
identified as {\it Bose Einstein condensation} (BEC) at
certain temperature ($T_{BEC}$) [5, 13] and the LT phase
should have non-zero $n_{p=0}(T)$, -increasing smoothly
from $n_{p=0}(T_{BEC}) = 0$ to $n_{p=0}(T = 0) = 1.0$
through $n_{p=0}(T < T_{\lambda}) < 1.0$.  The idea was
initially criticized by Landau [14] because LHE-4 is not a
SNIB.  But the Bogoliubov's theory of weakly interacting
bosons [15], concluding that the increasing strength of
interactions {\it simply} depletes $n_{p=0}(T)$, provided
some ground.  Consequently, different mathematical models
such as Bogoliubov prescription [15], pseudo-potential
technique [16], Jastrow formalism [17, 18], Feenberg's
perturbation [2, 18], {\it etc.} have been used to find the
magnitude of $n_{p=0}(T)$ in $He-II$ and at the same time
develop a general framework of microscopic theories
hereafter known as {\it conventional theories} (CTs).
While several articles published in [10] discuss the role
of BEC to understand the LT behavior of widely different
many body quantum systems, papers by Moroni {\it et. al.}
[19] and Sokol [20] make useful conclusions in relation
to our conventional understanding of LHE-4.  It appears
that CTs follow two different approaches, {\bf A1} and
{\bf A2}.

\bigskip
\noindent 
{\bf A1 :} This approach developed by Bogoliubov [15],
Beliaev [21, 22], Hugenholtz and Pines [23], Lee and Yang
[24], Brueckner and Sawada [25], Wu [26], Sawada [27],
Gavoret and Nozieres [28], and Hohenberg and Martin [29],
DeDominicis and Martin [30] has been reviewed by Woods and
Cowley [4] and Toyoda [31].  While it is elegantly
introduced by Fetter and Wadecka [32], its important
aspects are summed up, recently, by Nozieres [33], and Huang
[34].  A theoretical formulation following this approach
starts with the hamiltonian written in terms of second
quantized Schr\"{o}dinger fields and proceeds by using
important inferences such as: (i) the {\it hard core} (HC)
potential can be used perturbatively by using a method such
as pseudopotential method [16], (ii) coupling constant can
be related directly to the two body {\it scattering length}
($\sigma$) [22], (iii) a dimensionless expansion parameter
($n{\sigma}^3$) can be used to make perturbative calculation
[31], and (iv) the "depletion" of $n_{p=0}(T)$ can be
treated in the conventional perturbation by introducing
$c$-number operator for $p=0$ bosons [35].  Using the
quantum statistical mechanical expectation value of the
second quantized field operator as an {\it order
parameter} (OP) of the transition [23, 29], it then
calculates the relevant part of {\it free energy} as a
function of this OP and explains superfluidity and
related properties.  To this effect, one uses
sophisticated mathematical tools as discussed in
[31, 36].

\bigskip
\noindent
{\bf A2 :} This approach aims at finding the {\it radial
distribution function} [$g(r)$] and liquid structure
factor [$S(Q)$] which can be used to calculate the {\it
ground state} (G-state) properties, excitation spectrum,
different thermodynamic properties and equation of state.
Superfluidity and related properties are explained in
terms of one body density matrix $\rho{(r)}$ whose
asymptotic value at large $r$ gives the condensate
fraction $n_{p=0}(T)$ [37].  While different aspects of
this approach and related subject are discussed elegantly
by Feenberg [2], Croxton [38], Ceperley and Kalos [39]
and Yang [37] and its application to LHE-4 has been
reviewed by Smith {\it et. al.} [40(a)], Campbell and
Pinski [40(b)], Schmidt and Pandharipande [40(c)], Reatto
[40(d)], and Ristig and Lam [40(e)], several methods of
computer calculation of $g(r)$, $S(Q)$ and related
properties are discussed and reviewed in [41-43].  The
approach has been used by Moroni {\it et. al.} [19] and
Kallio and Piilo [44] to determine the properties of
LHE-4 and electron gas, respectively.  While a
comprehensive list of important papers related to this
approach are also available in [19, 44], developments
in our conventional understanding of quantum fluids
can be seen in most recent books and reviews [45-48].

\bigskip
Over the last seven decades a large number of papers have
been published [49] to establish the existence of
$n_{p=0}(T)$ in $He-II$ and to find its value.  However,
several prominent scientists working in the field have
expressed their doubts about the success of these
approaches in concluding the desired theory.  For example,
reviewing the progress of theoretical work at the
fifteenth Scottish university summer school (1974),
Rickayzen [50] states,``There is no microscopic
theory of superfluid $^4He$.  There are many mathematical
models which appear to provide insight into the behavior
of superfluids but there is no theory which provides
quantitative prediction that agree with observation.  Even
some of the most widely held assumptions of the theory such
as the idea of condensation in a zero momentum state can
not be said to be proved beyond reasonable doubt."  In
their review on the subject, Woods and Cowley [4] also
observe, ``Despite all the experimental information and
the numerous theoretical discussions there is still no
convincing theory of the excitations which begins with the
known interaction between helium atoms."  Interestingly,
though Kleban [51] proved that theories assuming the
existence of $p=0$ condensate contradict the {\it excluded
volume condition} (EVC),- a direct consequence of HC
nature of particles but his inference did not receive due
importance.  Similarly, Sokol [20] makes interesting
observations which doubt the accuracy of the inference
about the existence of $p=0$ condensate in $He-II$.  He
observes that: (i) neutron inelastic scattering
experiments have so far not given any indication of the
direct evidence ({\it i.e.} a $\delta$-function peak in
the momentum distribution ($n_p$) at $p=0$) of the
existence of $n_{p=0}(T)$ in $He-II$ and it is unlikely
that this goal will ever be reached, and (ii) in the
absence of detailed microscopic theory the estimate of
$n_{p=0}(0)$ $\approx$ 0.1 from different experiments
depends on current theories, models, and empirical
relations used to interpret experimental results and if
their underlying assumptions are incorrect, such
estimates of $n_{p=0}(T)$ will have no meaning.  In
addition though significant progress has been made in
refining our mathematical methods and computational
techniques applied to many body systems (including
liquid $^4He$) as one finds from most recent books and
reviews [45-48] but the presumed existence of $p=0$
condensate remains central to our present understanding
of the superfluidity of $He-II$ and similar SIB(s);
this is also evident from the most recent books [52]
summarizing the important aspects of LHE-4.

\bigskip
As such one finds that the best possible model that
could be  used to understand superfluidity is the
$\Psi-$theory [53] (a union of the two fluid model of
Landau [14], quantized circulation [54], vanishing of
superfluid density at the boundaries of the system [5],
{\it etc.}) which has emerged as exceedingly successful
phenomenology.  However, the microscopic theory of a
phenomenon has greater significance because only its
agreement with experiments testifies the validity of the
basic theory of nature, the above mentioned observations
motivated us to look for an alternative approach.
To this effect our initial reports [55-58] projecting the
basic aspects of our approach greatly helped us in
crystallizing our scientific ideas, refining our physical
arguments and mathematical formulations and laying down
the basic foundations for the final form of our theory
that we report in this paper.

\bigskip
Our theory is based on the macro-orbital representation of
a particle which we concluded in our recent paper [59]
related to the wave mechanics of two HC particles in 1-D
box and used in our studies related to the unification of
the physics of fermionic and bosonic systems [60], ground
stated of $N$ HC core particles in 1-D box [61] and basic
foundations of the microscopic theory of superconductivity
[62].  Our theory makes no assumption about the basic
factors ({\it viz.} the existence of $p=0$-condensate or
bound pair formation) which could, presumably, be
responsible for the superfluidity of a SIB.  In stead
we only use the solutions of our $N-$body Schr\"{o}dinger
equation to reach our conclusions.  In fact our theory,
for the first time, use the results of an obvious wave
mechanical superposition of two particles to conclude the
orderly placement of particles in phase space of the
system and this simple effort helped us in finding an
almost {\it exact theory} of a SIB which explains all
unique properties of $He-II$.  In variance with CTs, our
theory concludes that the origin of superfluidity and
related properties of a SIB lies with the BEC condensation
of $({\bf q}, -{\bf q})$ bound pairs in a state of their
center of mass (CM) momentum ${\bf K}=0$.

\bigskip
The paper has been arranged as follows.  Analyzing the 
$N$-particle microscopic quantum hamiltonian, Section (2.0)
identifies a pair of particles as the basic unit of the
system, uses macro-orbital representation (Appendix -A) of a
particle in writing $N-$particle state function and concludes
G-state configurations of a SIB and related aspects.
While Section (3.0) analyzes the thermodynamic behavior of the
system, Section (4.0) reports two different approaches used
to study thermal excitations.  While the origin of ({\bf q},
-{\bf q}) bound pair formation, energy gap, {\it etc.}, are
studied in Section (5.0), the consequences of the gap are
examined in Section (6.0) and the consistency of our theory
with: (i) phenomenological theories and (ii) experimental
results on LHE-4 is analyzed in Section 7.0.  Finally, we
present some important concluding remarks in Section 8.0

\bigskip

\centerline {\bf 2.0  Basic Foundations of Our Theory}

\bigskip
\noindent
{\bf 2.1 Hamiltonian}

\bigskip
A system of $N$ interacting bosons such as liquid $^4He$
can be described, to a good approximation, by
$$H(N) = \sum_i^Nh_i + \sum_{i<j}V(r_{ij}) \quad \quad
 \quad {\rm with } \quad
h_i = -({\hbar^2}/{2m})\bigtriangledown_i^2,  \eqno(1) $$

\noindent
where $m$ is the mass of a particle and $V(r_{ij})$ is a
two body {\it central force potential} which can be
expressed as the sum of: (i) a short range strong repulsion
$V^R(r_{ij})$ and (ii) a weak attraction $V^A(r_{ij})$
of slightly longer range.  To a good approximation, while
$V^R(r_{ij})$ can be equated to {\it hard core} (HC)
interaction $V_{HC}(r_{ij})$ [defined by $V_{HC}(r_{ij}
< \sigma) = \infty$ and $V_{HC}(r_{ij} \ge \sigma) = 0$
with $\sigma$ being the HC diameter of a particle],
$V^A(r_{ij})$ can be replaced by a constant negative
potential, {\it say} $-V_o$.  Although, this seems to
imply that the main role of $V^A(r_{ij})$ is to keep
particles within volume (V) of the system but as
concluded in Section (5.0) it also plays an important role
in binding two HC bosons having equal and opposite
({\bf q}, -{\bf q}) momenta.  However, to begin with,
our system can be identified as an ensemble of $N$ HC
bosons confined to a volume V and its  hamiltonion
can be written as,
$$H_o(N) = \sum_i^Nh_i + \sum_{i>j}^NA\delta{(r_{ij})}. 
\eqno(2) $$

\noindent
where we use
$$V_{HC}(r_{ij}) \equiv A\delta{(r_{ij})},  \eqno(3) $$

\noindent
with $A$ being the strength of Dirac delta repulsion.  For
impenetrable HC particles, we have $A \to \infty$ when
$r_{ij} \to 0$.  While the equivalence
expressed by Eqn.3 is mathematically shown by Huang [63],
its physical basis can be understood by examining the
possible configuration of two HC particle ({\it say}, P1
and P2) right at the instant of their collision.  When P1
and P2 during a collision have their individual CM located,
respectively, at ${\rm r}_{CM}(1) = \sigma/2$ and
${\rm r}_{CM}(2) = -\sigma/2$ (with ${\rm r}_{CM}$ being
the distance of the CM of a particle from the CM of the
pair of P1 and P2), they register their physical touch at
$r = 0$ and their encounter with $V_{HC}(r_{ij})$ is a result
of this touch beyond which the two can not be pushed in.  The
process of collision only identifies this touch; it does not
register how far are the CM points of individual particles
at this instant. In other words the rise and fall of the
potential energy of P1 and P2 during their collision at
$r = 0$ is independent of their $\sigma$ and this justifies
$V_{HC}(r_{ij}) \equiv A\delta{(r_{ij})}$.  It may, however,
be mentioned that this equivalence will not be valid in
accounting for certain physical aspects of the system
({\it e.g.}, the volume occupied by a given number of
particles) where the real size of the particle assumes
importance.

\bigskip
\noindent
{\bf 2.2 Basic unit of the system}

\bigskip
In what follows from Section 2.1, the motion of each
particle in the system can be described by a plane wave
$$u_{\bf p}({\bf b} ) = {\rm A}\exp(i{\bf p}.{\bf b})
\eqno(4)$$

\noindent
unless it collides with other particle(s).  Here A is
normalization constant, while {\bf p} and {\bf b},
respectively, are momentum ({\it wave number}) and position
vectors of the particle.  However, the plane wave
description is modified when the particle collides with
its neighbors or boundary walls of the system.  A
collision of a particle could
either be a two body collision or a many body collision
({\it e.g.}, two mutually colliding particles also collide
simultaneously with other particle(s)).  In the former case
two colliding particles (P1 and P2) simply exchange their
momenta ${\bf p}_1$ and ${\bf p}_2$ or positions ${\bf b}_1$
and ${\bf b}_2$ without any difference in the sum of their
pre- and post-collision energies.  However, in the latter
case P1 and P2 could be seen to jump from their state of
${\bf p}_1$ and ${\bf p}_2$ to that of different momenta
${\bf p}'_1$ and ${\bf p}'_2$ (possibly of different
energy) but it is clear that to a good approximation they
remain in one of the possible states of two HC particles
moving in the absence of other particle(s).  Evidently, the
complex dynamics of the system can be described, to a good
approximation, in terms of the simple dynamics of a pair of
HC particles as its {\it basic unit} and this is consistent
with the fact that inter-particle forces in our system are
basically two body forces.  We analyze the wave mechanics
of a pair of HC particles in Appendix-A and use its
inferences in the following analysis.  

\bigskip
\noindent
{\bf 2.3 State function}

\bigskip
As concluded in Appendix-A, each particle in our system is
represented more accurately by a macro-orbital ({\it cf.}
Eqn. A-17) rather than by $u_{\bf p}({\bf b})$. Accordingly,
each particle is a member of the pair of particles moving
with equal and opposite momenta ({\bf q}, - {\bf q}) at
their CM which moves with moment {\bf K} in the laboratory
frame.  It has two motions, $q$ and $K$.  While $q$ defines
its quantum size ($\lambda/2 = \pi/q$) representing
the size of the real space occupied exclusively by it, $K$
defines its motion as a free particle of mass $4m$.  By
using $N$ macro-orbitals for $N$ particles, we obtain 
$$\Psi^j_n(N) = \Pi_i^N\zeta_{q_i}(r_i)\sum_P^{N!}
[(\pm 1)^P\Pi_i^N\exp{i(P{\bf K}_i.{\bf R}_i)}]  \eqno(5)$$

\noindent
as the state function which represents one of the $N!$
microstates of the system.  Here we have $\zeta_{q_i}(r_i) =
\sin{({\bf q}_i.{\bf r}_i)}$ and $\sum_P^{N!}$ referring to
the sum of $N!$ product terms obtainable by permuting $N$
particles on different ${\bf K}_i$ states with $(+1)^P$ and
$(-1)^P$, respectively, used for selecting a symmetric and
anti-symmetric wave function for an exchange of two
particles.  In principle, the permutation of $N$ particles
on different ${\bf q}_i$ states also gives $N!$ different
$\Psi^j_n(N)$ and this renders
$$\Phi_n(N) = \frac{1}{\sqrt{N!}}\sum_j^{N!}\Psi^j_n(N)
\eqno(6)$$

\noindent
as the complete wave function of a possible quantum state
of the system.

\bigskip
\noindent
{\bf 2.4 State energy}

\bigskip
Since each macro-orbital representing a particle (say $i-$th)
in $\Phi_n(N)$ state has a reference to the CM coordinate
system related to the pair of which the $i-$th particle is a
member, we need to recast $H_o(N)$ (Eqn. 2) in terms of these
coordinates in order to examine if $\Phi_n(N)$ is an
eigenfunction of $H_o(N)$.  To this effect, we define
$$h(i) = \frac{1}{2}[h_i + h_{i+1}] =
-\frac{\hbar^2}{8m}\bigtriangledown_{R_i}^2
-\frac{\hbar^2}{2m}\bigtriangledown_{r_i}^2
 \eqno(7)$$

\noindent
${\rm with} \quad  h_{N+1} = h_1$ and write Eqn. 2 as 
$$H_o(N) =  \sum_i^Nh(i) + \sum_{i>j}^NA\delta{(r_{ij})}.
\eqno(8)$$

\noindent
This easily renders [58-62]  
$$<\Phi_n(N)|\sum_{i>j}^NA\delta{(r_{ij})}|\Phi_n(N)> = 0
\eqno(9)$$

\noindent
and  
$$E_n =<\Phi_n(N)|H_o(N)|\Phi_n(N)> =
\sum_i^N\left[\frac{\hbar^2q_i^2}{2m}  +
\frac{\hbar^2K_i^2}{8m}\right]  \eqno(10)$$

\noindent
which concludes that $\Phi_n(N)$ is an eigenstate of
$H_o(N)$ with $E_n$ as its energy eigenvalue.  While, at
the first glance, Eqn. 10 indicates that $E_n$ is purely a
kinetic energy but the fact, that $q$ values are constrained
to satisfy $q \ge q_o = \pi/d$ ({\it cf.} Appendix-A, Eqn.
A-13), indicates that HC
interaction has an obvious control on $E_n$; of course this
impact comes to surface only at low temperatures at which
quantum effects start dominating the behavior of the system.

\newpage
\bigskip
\noindent
{\bf 2.5  G-state configuration} 

\bigskip 
In the light of our inference that $q \ge \pi/d$
({\it cf.} Appendix-A, Eqn. A-13) and $K$ can have any
value between 0 and $\infty$, the G-state energy of a
SIB can be obtained by using all $q_i = \pi/d_i$ and all
$K_i =0$ in Eqn. 10 and this renders 
$$E_o = \sum_i^N \frac{h^2}{8m{d_i^2}}
= \sum_i^N \frac{h^2}{8m{{\rm v}_i^{2/3}}}
\quad \quad {\rm with}  \quad \sum_i^N
{\rm v}_i = {\rm V} \quad {\rm (constant)}   \eqno (11)$$

\noindent
where we use $d_i = {\rm v}_i^{1/3}$ with ${\rm v}_i$ being
the volume of the real space,- exclusively occupied by the
$i-$th particle.  In writing $\sum_i^N{\rm v}_i = {\rm V}$,
we use the fact that each particle of the lowest possible
$q$ has largest possible quantum size $\lambda/2$ and,
therefore, occupy largest possible ${\rm v}_i$.
While different particles can, in principle, occupy
different volumes but the simple algebra reveals that $E_o$
is minimum value for ${\rm v}_1 = {\rm v}_2 =..{\rm v}_N
= {\rm V}/N$ and we have  
$$E_o = Nh^2/8md^2 = N{\varepsilon}_o.    \eqno (12)$$

\noindent
Since this means that all particles, identically, have
$q = q_o$ and $K = 0$, use of these values in Eqn. 6
renders
$$\Phi_o(N) = \Pi_i^N\zeta_{q_o}(r_i) =
\Pi_i^N\sin(q_or_i) \eqno(13)$$

\noindent
as the G-state wavefunction.  It is clear from
Eqns. 12 and 13 that each particle in the system
represents a particle trapped in a box (cavity formed by
its neighboring particle) of size $d$  and it rests at
the central point of this cavity.  Since each
$\sin(q_or_i)$ in Eqn. 13 represents a kind of
stationary matter wave (SMW) which joins with other SMWs
of neighboring particles at the boundaries between the
two cavities they occupy, the G-state
wave function seems to be a macroscopically large size
3-D network of SMWs which modulates the {\it relative
positions} of two particles in phase space and extends
from one end of the container to another.  In Section (5.0),
we find that this network gets energetically stabilized due
to some kind of collective binding and assumes different
aspects of macroscopic wave
function of S-state as envisaged by London [12].  
Using this information with our results [Eqns. A-13 and
A-14 of Appendix-A)] related to $<r>$ and $<\phi>$ (the
expectation of relative position of two particles in $r-$
and $\phi-$spaces), we find that the G-state
configuration of particles can be described by
$$<k> = 0 \quad\quad <r> = d \quad\quad {\rm and}
\quad\quad<\phi> = 2n\pi \quad {\rm with} \quad n=1,2,3 ...
\eqno(14)$$

\noindent
Evidently, particles in the G-state cease to have
relative motion, collisional motion and inter-particle
scattering.  Since all of them
have identically equal $<r> = d$ (nearest neighbor distance)
with constant distance of $2\pi$ in the $\phi-$space, the
G-state represents their close packed arrangement in
real space and they are constrained to move only in order of
their locations.  In the light of these inferences and our
conclusion that particles in this state have a kind of
collective binding at $T \le T_{\lambda}$ ({\it cf.}
Section 5.0), it is evident that particles in the LT phase
have mutual binding in all the three spaces ($r-$, $q-$ and
$\phi-$).

\bigskip 
\noindent
{\bf 2.6  Evolution of the system on cooling} 

\bigskip 
In view of the condition, $\lambda/2 \le d$ ({\it cf.}
Appendix-A, Eqn. A-13), ($d - \lambda/2$) for a SIB of fixed
number density decreases with decreasing $T$ and at certain
$T = T_c$ it vanishes at large.  Consequently, the system at
$T_c$ has $q = q_o$ for all particles and its state (Eqn. 6)
is now expressed by
$$\Phi_n^S(N) =  \Phi_o(N)\sum_P^{N!}
[(\pm 1)^P\Pi_i^N\exp{i(P{\bf K}_i.{\bf R}_i)}]  \eqno(15)$$

\noindent
For the reasons which become clear in Sections 5.0 and 6.0,
the superscript $S$ in $\Phi_n^S(N)$ refers to S-phase.
Eqn. 15 implies that all the $N!$ micro-states of
our system appearing in $\Phi_n(N)$ (Eqn. 6) merge into one
and the entire system at $T_c$ attains a kind of oneness as
envisaged by Taubes [64].  This also reveals that: (i) the
system at $T_c$ has two separate components, say F1 and F2.
While F1 (described by $\sum_P^{N!}..$ part of $\Phi_n^S(N)$)
represents a gas of non-interacting quasi- particle
excitations originating from the plane wave $K-$motions
of particles, F2 (described by $\Phi_o(N)$) represents the
system in its G-state (or $T=0$ state) which has no
motion except zero-point $q-$motions.  The density of F1,
obviously, decreases with cooling the system and reaches
zero value at $T = 0$.  In Section-7.0, this separation
is used to explain the two fluid behavior of Phase-II.  

\bigskip
\noindent
{\bf 2.7  Quantum Correlations} 

\bigskip
Quantum correlations, basically, originating from the wave
nature of particles play an important role in relation to
the behavior of a quantum system.  These correlations can be
expressed in terms what is known as {\it quantum correlation
potential} (QCP) which can be obtained
[65, 66] by comparing the partition function (under the
quantum limits of the system), $Z_{\rm q} = {\sum}_n
\exp(-E_n/k_BT)|{\Phi}_n(S)|^2$ and its classical equivalent,
$Z_{\rm c} = {\sum}_n \exp(-E_n/k_BT)\exp(-U_n/k_BT)$.  Here
${\Phi}_n(S)$ is given by Eqn. 15.  The application of this
method to our system is justified because our theory describes the
system by symmetrized plane waves rendering $<V_{HC}(r)> = 0$
which indicates that the HC potential is completely screened out.
Simplifying $U_n$, one easily finds that
pairwise QCP has two components.  The $U^s_{ij}$ pertaining to
$k$ motion controls the $\phi = kr$ position of a particle and
we have
$$U^s_{ij} = - k_BT_o\ln[2\sin^2({\phi}/2)]  \eqno (16)$$ 

\noindent
where $T$ has been replaced by $T_o$ because $T$ equivalent
of $k$ motion energy at all $T \le T_{\lambda}$ is $T_o$.      
 
\bigskip
$U^s_{ij}$ has minimum value $(-k_BT_o\ln{2})$ at
${\phi} = (2n+1){\pi}$ and maximum value $(=\infty)$ at
${\phi} = 2n{\pi}$ occurring periodically at $\Delta\phi
= 2n{\pi}$ (with $n = 1,2,3,...$).  Since $U^s_{ij}$ always
increases for any small change $\delta\phi$ in $\phi$
at its minimum value, as
$${\frac{1}{2}}{\rm C}({\delta\phi})^2 =
{\frac{1}{4}}k_BT_o({\delta\phi})^2 \quad \quad
{\rm with \,\, force \,\, constant}  \quad \quad
{\rm C} = {\frac{1}{2}}k_BT_o \eqno (17) $$

\noindent
and particles experience a force = $-$ C${\delta\phi}$
which tries to maintain $\delta\phi = 0$ and the order of
particles in $\phi$-space is sustained.  Since $U^s_{ij}$
is not a real interaction like $V(r)$ which can manipulate
$d$, the $\phi$-space order is, therefore, achieved when
cooling drives all $q$ towards $q_o$.

\bigskip
The second component pertaining to $K-$motions can be
expressed by [56 and 65]
$$U_{ij} = -k_BT\ln{[1 +
\exp{(-2\pi|R'-R''|^2/\lambda_T'^2)}]} \eqno (18)$$

\noindent
with $\lambda_T'= h/\sqrt{2\pi (4m)k_BT}$ being the thermal
de Broglie wavelength pertaining to $K-$motions for which
each particle appears to have $4m$ mass.  Note that $U_{ij}$
is identical to the quantum correlation potential for
non-interacting particles [65].  It may be seen as the origin
of the force that facilitates BEC of particles at the $K = 0$
state by driving them towards $K = 0$ point in $K-$space where
$U_{ij}$ has its minimum value $-k_BT\ln 2$.

\bigskip
\centerline {\bf 3.0  Thermodynamic Behavior}

\bigskip
\noindent
{\bf 3.1  Equation of state}

\bigskip
In what follows from Eqn. 10, we can express the energy
of a particle in our system as 
$$\epsilon = \varepsilon{(K)} + \varepsilon{(k)} = 
\frac{{\hbar}^2K^2}{8m} +
\frac{{\hbar}^2k^2}{8m} \eqno(19)$$

\noindent
which can have any value between ${\varepsilon}_o$ and
$\infty$,  Interestingly, this possibility exists even if
${{\hbar}^2k^2}/{8m}$ is replaced by the lowest energy
${\varepsilon}_o$ of its $q-$motion since $K$ can have
any value between 0 and $\infty$ and we can use 
$$\epsilon = \frac{{\hbar}^2K^2}{8m} + {\varepsilon}_o
\eqno(20)$$

\noindent
in the starting expressions of the standard theory of BEC
[67, Ch. 7] to obtain
$$\frac{PV}{k_BT} = -{\Sigma}_{\varepsilon{(K)}}
\ln{[1-z\exp{(- \beta[{\varepsilon{(K)} +
{\varepsilon}_o}]})}] \eqno (21)$$

\noindent
and
$$N = {\Sigma}_{\varepsilon{(K)}}\frac{1}{z^{-1}\exp{(\beta
[{\varepsilon{(K)} + {\varepsilon}_o}])} -1}  \eqno (22)$$

\noindent
with $\beta = \frac{1}{k_BT}$ and fugacity 
$$z=\exp{({\beta}{\mu})} \quad \quad (\mu = {\rm chemical
\quad potential}). \eqno(23)$$

\noindent
Once again, by following the steps of the standard theory of
BEC [67] and redefining the fugacity by 
$$z' = z\exp{(-\beta{\varepsilon}_o)} =
\exp{[\beta(\mu - {\varepsilon}_o)]} =
\exp{[\beta\mu']} \quad \quad {\rm with} \quad
\mu' = \mu - {\varepsilon}_o  \eqno(24)$$

\noindent
we easily have  
$$\frac{P}{k_BT} = - {\frac{2\pi{(8mk_BT)^{3/2}}}{h^3}}
{\int}_0^{\infty}x^{1/2}\ln{(1-z'e^{-x})}dx =
\frac{1}{{\lambda}^3}g_{5/2}(z') \eqno(25) $$

\noindent
and 
$$\frac{N-N_o}{V} = {\frac{2\pi{(8mk_BT)^{3/2}}}{h^3}}
{\int}_0^{\infty}\frac{x^{1/2}dx}{z'^{-1}e^x -1} =
\frac{1}{{\lambda}^3}g_{3/2}(z') \eqno(26)$$

\noindent
where $x = \beta{\varepsilon}(K)$, $\lambda =
h/(2{\pi}(4m)k_BT)^{1/2}$ and $g_n(z')$ has its usual
expression.  This reduces our problem to that of non-
interacting bosons but with a difference.  Firstly, we have
$m$ replaced by $4m$ and $z$ by $z'$.  Secondly, the
theory of non-interacting bosons concludes $z = 1$
(or $\mu = 0$) for $T \le T_{\lambda}$ and $z < 1$
(or $\mu < 0$) for $T > T_{\lambda}$, while our theory of
interacting bosons fixes $z' = 1$ [or $\mu' = 0$
rendering $\mu = {\varepsilon}_o$ (Eqn. 24)] for $T \le
T_{\lambda}$ and $z' < 1$ (or $\mu' < 0$ demanding $\mu <
{\varepsilon}_o$) for $T > T_{\lambda}$.  In other words
we have $z'$ and $\mu'$ in place of $z$ or $\mu$ used in
the theory of non-interacting bosons [67].  As such we can
use Eqns. 21 and 22 and Eqns. 25 and 26 to evaluate
different thermodynamic properties.  For example, using
Eqns. 25 and 26, we find the internal energy
$U = - \frac{\partial}{\partial\beta}(\frac{PV}{k_BT})|_{z,V}$
of the system.  We have
$$U = \frac{3}{2}k_BT\frac{V}{{\lambda}^3}
g_{5/2}(z') + N\varepsilon_o = U' +
N\varepsilon_o  \eqno(27)$$

\noindent
where $U'= - \frac{\partial}{\partial\beta}
(\frac{PV}{k_BT})|_{z',V}$, obviously, represents internal
energy of $K-$motions, while $N\varepsilon_o$ comes from
$k-$motions.  Similarly, the Helmholtz free energy of the
system can be expressed as 
$$F = N\mu - PV = N\varepsilon_o + (N\mu'-PV)
= N\varepsilon_o + F' \eqno(28)$$

\noindent
with $F'$ referring to that of non-interacting bosons.
Following the standard methodology, we may now analyze
$F$ for the physical conditions for which it becomes
critical and leads to superfluidity.

\bigskip
\noindent
{\bf 3.2  Onset of $K=0$ condensate and $T_{\lambda}$}

\bigskip
In what follows from Eqn. 28, $F$ is the sum of: (i)
$F' = F(K)$, representing the contribution of
$K-$morions which define a system of non-interacting
quantum quasi-particles of ({\it bosonic nature}) and
$4m$ mass and (ii) $N\varepsilon_o = F(q)$ representing
the zero-point energy of $q-$motions.  Following
the standard theory of noninteracting bosons [67],
$F(K)$ is expected to become critical at
$$T_b  = \frac{1}{4}T_{\rm BEC} =
\frac{h^2}{8{\pi}mk_B}{\left({\frac{N}
{2.61{\rm V}}}\right)}^{\frac{2}{3}}
\eqno (29)$$  

\noindent
with the onset of BEC of particles in their $K=0$
state.  $T_{\rm BEC}$ in Eqn. 29 represents the usual
point of BEC in a system of non-interacting bosons and
$\frac{1}{4}$ factor signifies that each boson for its
$K-$motion behaves like a particle of mass $4m$ and 
$T_{\rm BEC}$ varies as $\frac{1}{m}$.  However, this 
onset would occur only when the particles at large have 
fallen into the G-state of the other component of their 
motion ({\it viz.} $q-$motions) because in an otherwise 
situation $K-$motions can gain energy from $q-$motion 
energy.  In what follows from Section 2.6, a SIB assumes 
a state of $q= q_o$ for all particles at
$$T_o =  \frac{h^2}{8{\pi}mk_B}\frac{1}{ d^2}
\eqno (30)$$

\noindent
which represents the $T$ equivalent of the G-state energy
$\varepsilon_o$.  It is evident that as soon as all
particles have $q=q_o$ state, further cooling of the
system would immediately trigger the process of their fall
in $K=0$ state.  Evidently, the effective temperature, at
which particles have the state of $q = q_o$ and the onset of
$K=0$ condensate, should be
\bigskip
$$T_{\lambda} =
 \frac{h^2}{8{\pi}m^*k_B}{\left[{\frac{1}{ d^2}} +
{\left({\frac{N}{2.61{\rm V}}}\right)}^{\frac{2}{3}}\right]}
  \eqno (31)$$  

\bigskip
\noindent
Here we replace $m$ by $m^*$ to account for the impact of
physical parameters such as pressure ($P$) on $T_{\lambda}$.
In this context it may be noted that, in spite of the fact
that particles in our system to a good approximation
represent HC particles moving freely on a surface of constant
$-V_o$, one can not ignore the role of inter-particle
interaction during their relative motion.  We note that $d$,
V and $m^*$ are three quantities which may change with
increasing $P$.  While $T_{\lambda}$ is expected
to increase with increasing $P$ for the usual decrease in
the values of $d$ and V, however, $T_{\lambda}$ may show a
reverse change if $m^*$ increases with $P$.  In this
context we note that $m^*$ for $^4He$ atoms in LHE-4
should increase with $P$ for an obvious increase in the
strength of inter-particle {\it attraction} with increasing
$P$.  Evidently, Eqn. 31 can explain the $P$ dependence
of $T_{\lambda}$ of LHE-4.

\bigskip
\noindent
{\bf 3.3  Nature of Transition}

\bigskip
When our system is cooled through $T_{\lambda}$, its
particles move from their state of $q > \pi/d$
({\it i.e.}, $\phi > 2\pi$) to that of $q = \pi/d$
({\it i.e.}, $\phi = 2\pi$) ({\it cf.} Section-2.5).
Evidently the system transforms from a state of random
distribution of its particles in $\phi-$space to that of
orderly distribution with $\phi = 2n\pi$ ($n=1,2,3,$ ..).
This means that $\lambda-$transition represents an onset of
order-disorder of particles in $\phi-$space accompanied
by the BEC of particles in the state of $K=0$ and
$q=\pi/d$.  Since BEC of particles in $K=0$ state is not
different from the BEC of non-interacting bosons (a well
known second order transition [67]) and their order-disorder
basically represents their self-organization in $\phi-$space
which results from a reshuffle of their momenta,
it is evident that $\lambda-$transition is a second order
transition.

\newpage
\bigskip
\noindent
{\bf 3.4 Free energy and order parameter of $T_{\lambda}$}

\bigskip
In the light of Section 3.1, it is evident that free energy
$F$ of the system has two components and it can be
expressed as
$$F = F(q) + F(K) \approx N\varepsilon_o + F(K) \eqno (32)$$

\noindent
with 
$$F(K) = k_BT.{\frac{2\pi{(8mk_BT)^{3/2}}}{h^3}}
 {\int}_0^{\infty}x^{1/2}\ln{(1-z'e^{-x})}dx
 = k_BT\frac{1}{{\lambda}^3}g_{5/2}(z') \eqno (33)$$

\noindent
Eqn. 32 would also hold for $T > T_{\lambda}$, provided
such a $T$ is low enough to allow only insignificantly low
number of particles in the excited states of momentum
$q \ge 2q_o$.  Guided by the experimental fact that a
system like LHE-4 exhibit superfluidity and related
properties below $T_{\lambda}$, $F$ can be expressed as 
$$F(T, \Omega) = F_o + \frac{1}{2}A\Omega^2 +
\frac{1}{4}B\Omega^4 + \frac{1}{6}C\Omega^6 +
...  ...  \ \eqno (34)$$

\noindent
with $\Omega$ being the order parameter, $F_o$ being the
constant component of $F$ and $A$, $B$, $C$, $etc.$ being
the coefficients of its expansion which may depend on physical
conditions such as $T$ and $P$.  However, it may be mentioned
that it is $F(q)$ (not $F(K)$) which is basically responsible
for the unique properties of a SIB below $T_{\lambda}$ since
a gas of non-interacting bosons is not expected to exhibit
superfluidity and related properties in spite of its BEC into
zero momentum state.
Defining $A = \alpha.(T-T_{\lambda})/T_{\lambda}$, we find that
$$\Omega{(T)} \propto \sqrt{\frac{T_{\lambda}-T}{T_{\lambda}}}. 
\eqno (35)$$

\noindent
However, since $T_o$ represents the $T$ equivalent of the
G-state energy of a particle $\varepsilon_o$, there is
hardly any excitation in the system at $T_o$.  Consequently,
the state of the system at $T_o$ is effectively a $T = 0$
state and for this reason we can renormalize the temperature
scale by replacing $T$ in Eqn. 35 by $T^* = T -T_o$ and
recast $\Omega{(T)}$ as
$$\Omega{(T^*)} \propto \sqrt{\frac{T^*_{\lambda}-T^*}
{T^*_{\lambda}}} = \sqrt{\frac{T_{\lambda}-T}
{T_{\lambda} - T_o}}  \eqno (36)$$

\noindent
which rightly indicates that $\Omega{(T^*)}$ reaches
its maximum value 1.0 at $T = T_o$.  

\bigskip
Now we may identify the physical nature of $\Omega$ and
to this effect our first choice is the fraction of the
number of bosons ($n_{K=0}(T) = N_{K=0}(T)/N$) condensed
into the G-state defined by $q = q_o$ and $K=0$.
Applying the standard theory of BEC of non-interacting
bosons [67] to F1, we find that $n_{K=0}(T)$ should vary as
$[1 - (T/T_{\lambda})^{3/2}]$.  However, for our
system $n_{K=0}(T)$ is replaced by $n_{K=0}(T^*)$ which
varies as
$$n_{K=0}(T^*) = \frac{N_{K=0}(T^*)}{N} =
\left[1 - \left(\frac{T^*}{T_{\lambda}^*}\right)^{3/2}\right] = 
\left[ 1 - \left(\frac{T-T_o}{T_{\lambda} - T_o}\right)^{3/2}\right]
\eqno (37)$$

\noindent
Our next choices could be: (i) the strain in the
interparticle bonds [$(d_T - d_{\lambda})/d_{\lambda}$] whose
origin, relation with $({\bf q}, - {\bf q})$ bound pair
formation, volume  expansion on cooling through $T_{\lambda}$,
and energy gap has been discussed at length in Section-5.0,
and (ii) $n^* = [N^*(T_{\lambda})- N^*(T)]
/N^*(T_{\lambda})$ which represents the fraction of particles
that move from their state of disorder in $\phi-$space to
that of order on cooling though $T_{\lambda}$ ({\it cf.}
Eqn. 58 for an expression for $N^*(T)$).  We
analyze $n^*$ and its relation with the energy gap
in Section-5.0.  One may find that all these quantities are
inter-related with each other.  

\bigskip
\noindent
{\bf 3.5  Single particle density matrix and ODLRO}

\bigskip
Using Eqn. 15, we obtain the single particle density matrix,
$$\rho({\bf R^*-R}) = \frac{N}{\rm V}\left[\frac{N_{K=0}(T^*)}{N} + 
\frac{\rm V}{{\lambda}'^3_T}\exp{\left[- 2\pi.
\frac{|\bf {R^*-R}|^2}{{{\lambda}'_T }^2} \right]} \right]
\left({\frac{2}{\rm V}}
\sin^2[\frac{\pi(r''-r')}{d}]\right)  \quad \quad
   \eqno (38)$$

\noindent
with $N_{K=0}(T^*)/N = n_{K=0}(T^*)$ ({\it cf.}, Eqn. 37).
Here we use: (i) ${\bf q}_o.({\bf r''} -{\bf r'})^* =
2n{\pi} + {\bf q}_o.({\bf r''} -{\bf r'})$ and (ii) a
renormalized $T$ scale $T^* = T - T_o$.

\bigskip
While the term in big (..) of Eqn. 38 represents the
variation of density over a single antinodal region of the
SMW of a particle, $N_{K=0}(T^*)$ stands for the number of
particles condensed to the state of $K=0$ and $q = q_o$;
${\lambda}'_T = h/{\sqrt{2\pi(4m)k_BT}}$ represents thermal
wavelength attributed to $K$ motions.  We note that under
the limit  $|{\bf R^*} - {\bf R}| \to \infty$, the ``one
particle density matrix'' ($\rho({\bf R^*-R})$) has
nonzero value ($N_{K=0}(T)/V$) for $T<T_{\lambda}$ and
zero value for $T \ge T_{\lambda}$ since $N_{K=0}(T^*)$ is
$\ne 0$ for $T<T_{\lambda}$ and $=0$ for $T \ge T_{\lambda}$.
Evidently, our theory satisfies the criterion of Penrose and
Onsager [35] for the occurrence of BEC in the G-state of the
system defined by $K = 0$ and $q = \pi/d$ and agrees with
the idea of ODLRO, spontaneous symmetry breaking and phase
coherence advanced, respectively, by Yang [37], Goldstone
[68] and Anderson [69].

\bigskip
\noindent
{\bf 3.6  Logarithmic singularity of specific heat}

\bigskip
As shown in this section, this singularity is associated
with the process which reshuffles particles from their
random positions in $\phi-$space at $T^+_{\lambda}$ (just
above $T_{\lambda}$) to their ordered locations at
$T^-_{\lambda}$(just below $T_{\lambda}$).  We assume that
in this process about $N_{\lambda}$ particles move from
their ${\phi} = (2n{\pi} \pm \delta{\phi}_{\lambda})$
positions at $T^+_{\lambda}$ to ${\phi} = (2n+1){\pi}$ at
$T^-_{\lambda}$.  This releases $\Delta{\varepsilon}$
energy responsible for the singularity and we have
$$\Delta{\varepsilon} =
- N_{\lambda}k_BT_o\left[{(\ln{2}){\sin^2{\left(\frac
{2n{\pi} \pm \delta{\phi}_{\lambda}}{2}\right)}}
- \ln{2}}\right]  \eqno (39) $$    

\noindent
Following the theories [70] of critical phenomenon we may
define
$$\delta{\phi}_{\lambda}
= \delta{\phi}_{\lambda}(o)|\zeta|^{\nu}
\left[ 1 + a_2|\zeta|^2 + a_3|\zeta|^3
\right] \eqno (40)$$

\noindent
with $\zeta = \frac{T-T_{\lambda}}{T_{\lambda}}$.  
To a good approximation we have  
$$\Delta{\varepsilon} = -
N\left(\frac{T-T_{\lambda}}{T_{\lambda}}\right)
k_BT_o\ln{\left( \frac{\delta{\phi}_{\lambda}(o)
|\zeta|^{\nu}}{2}\right)^2}   \eqno (41)$$ 

\noindent
by using $\delta{\phi}_{\lambda} =
\delta{\phi}_{\lambda}(o)|\zeta|^{\nu}$ and 
$N_{\lambda} = N\frac{T-T_{\lambda}}{T_{\lambda}}$; the
latter expression is so chosen to ensure that
$\Delta{\varepsilon}$ does not diverge at $T_{\lambda}$
and it decreases with decreasing $T$ through $T_{\lambda}$.
Eqn. 41 gives
$$C_p(T \approx T_{\lambda})
\approx - \frac{N}{T_{\lambda}}k_BT_o[2\nu\ln|\zeta| 
+ \ln{(\delta{\phi}_{\lambda}(o)^2)}
- \ln{4} + 2{\nu}] \eqno (42)$$

\noindent
It may be mentioned that $\nu$, $\delta{\phi}_{\lambda}(o)$,
{\it etc.} may not have same values at $T > T_{\lambda}$
and at $T < T_{\lambda}$, obviously, because the relative
configuration of the system on the two sides of
$T_{\lambda}$ differ significantly.  

\bigskip
\centerline{\bf 4.0  Thermal Excitations.} 

\bigskip
\noindent
{\bf 4.1  Feynman's approach}

\bigskip
Defining an excited state of the system by $\psi =
\sum_if(r_i)\phi$ and the G-state by $\phi$, Feynman
[71] showed that the excited state energy is minimum
for $f(r_i) = \exp(i{\bf k}.{\bf r}_i)$ and he obtained
$$E(Q)_{Feyn} = \frac{{\hbar}^2{Q}^2}{2m{S(Q)}} \eqno(43)$$ 

\noindent
with $S(Q)$ = structure factor of the system.  However,
$E(Q)$ obtained from Eqn. 43 was found to be about two times
the experimental value.  Introducing back flow effect
Feynman and Cohen [72] later found better agreement with
experiments but with considerable discrepancy at higher $Q$.
In this section we use Feynman's approach to our particles
in ({\bf q}, -{\bf q}) configuration.  We note that under
the impact of an excitation a SMW pair in the G-state
configuration of (${\bf q}_o$, $-{\bf q}_o$) and $K=0$,
expressed by $\zeta (q_o)$ (Eqn. A-17, with $k = 2q_o$ and $K=0$)
moves to a new configuration (${\bf q}_o + \Delta{\bf q}$,
$-{\bf q}_o + \Delta{\bf q}$) described by
$$\zeta_{q_o}^*(r) = \zeta(q_o)(r)\exp{(i{\bf Q}.{\bf R})}
\exp{(-i\varepsilon{(Q)}t/\hbar} \eqno (44) $$ 

\noindent
with $2\Delta{\bf q} = {\bf Q}$ which means that the
impulse changes in the CM momentum and energy of both
particles and its impact should be identified by ${\bf Q}$
not by $\Delta{\bf q}$.  Eqn. 44 further shows that
$f(R) = \exp{(i{\bf Q}.{\bf R})}$ is the real form of $f$
which renders $\psi = \sum_i\exp{(i{\bf Q}.{\bf R}_i)}
\phi$.  Using these facts and recasting the relation,
$-(\hbar^2/2m)\nabla^2 f(R) = \varepsilon \int g(R-R')
f(R')d^3R'$ (Eqn.(11.25) of Feynman [71], p 329-30),
for $i$-th particle we find $-(\hbar^2/8m)\nabla^2_i f(R) =
\varepsilon_i(Q)S(Q)f(R)$; note that our single particle
energy operator is $-(\hbar^2/8m)\nabla^2_i$.  This renders
$(\hbar^2Q^2/8m) = \varepsilon_i(Q)S(Q)$.  Adding this
relation to a similar relation for $j$-th particle of the
pair gives
$$E(Q) = \varepsilon_i(Q) + \varepsilon_j(Q) =
\frac{{\hbar}^2{Q}^2}{4m{S(Q)}} = 
\frac{1}{2}E(Q)_{Feyn} \eqno (45)$$

\noindent
which naturally explains $E(Q)_{expt}$ of $He-II$.

\bigskip
\noindent
{\bf 4.2  Aproach of mono-atomic chain}

\bigskip
We note that a system like liquid ${^4He}$ is expected to
exhibit: (i) no transverse mode because the shear forces
between its particles are negligibly small, and (ii) only
one branch of longitudinal mode because the system is
isotropic.  Further since the particles in the G-state
of a SIB from a kind of close packed arrangement with
$\Delta\phi = kr = 2n{\pi}$ (Section 2.5), we can visualize
waves of $\phi$-oscillations. Using a linear chain of atoms
with nearest neighbor forces, we find that the frequency
dispersion of $\phi$-oscillations can be expressed by
$${\omega}_{\phi}{(Q)} =
{\sqrt {(4{\rm C})/{\beta}}}|\sin({Q}d/2)| \eqno (46)$$

\noindent
where $Q$ is the wave vector and $\beta$ is the measure of
inertia for ${\phi}$ motion.  Since $\phi-$positions of
particles are restored by $U^s_{ij}$, C is given by
Eqn. 17.  We note that $\phi$-oscillations can appear as
the oscillations of $r$ and $q$ because $\delta\phi =
2q{\delta}r + 2{\delta}qr$; we have phonons when $q$
remains unchanged at $q_o$, and omons (a new kind of quantum
quasi-particle representing a phononlike wave of the
{\bf o}scillations of {\bf m}omentum) coordinates when $r$
remains fixed at $d$.

\bigskip
Evidently, ${\omega}_r({Q})$ of phonons can be represented,
to a good approximation, by the dispersion of the elastic
waves in a chain of identical atoms and it can be obtained
from Eqn. 46 by replacing $\beta$ and C by $m$ and  $C^*$,
respectively.  We have
$${C^*} = 4{\pi}^2{\rm C}/{d}^2 = 2{\pi}^2k_BT_o/{d}^2
= {\pi}h^2/4md^4   \eqno (47)$$
 
\noindent
However, for a better accuracy $d$ and $C^*$ should,
respectively, be considered descending and ascending
functions of $Q$ because increase in the energy of
particles changed by an excitation reduces their WP size and
decreases $d$ and thereby increases $C$.  Consequently,
${E_{ph}(Q)}$ can be expressed more accurately by     
$${E_{ph}(Q)} = \hbar{\omega}_r({Q}) = 
 \hbar{\sqrt{4{C(Q)}/m}}|\sin({Q}d{(Q)}/2)|
 \eqno (48).$$

\noindent
We note that use of $C^*(Q)$ and $d(Q)$ not only explains
the experimentally observed ${E_{ph}(Q)}$ of $He-II$ but
also accounts for its anomalous nature at low $Q$
[56].  This aspect has been studied in detail in [73, 74].
However, since $d(Q)$ can not be
smaller than $\sigma$, $d(Q)$ and $C(Q)$ are bound to become
$Q$ independent for $Q > \pi/{\sigma}$ and the maximum in
${E_{ph}(Q)}$ ({\it i.e.} the position of so called maxon)
should fall at maxon $Q = {Q_{max}} = \pi/{\sigma}$ and
${E_{ph}(Q)}$ for $Q > \pi/{\sigma}$ and $ < 2\pi/d$ should
follow
$${E_{ph}(Q)} =  \hbar{\omega}_r({Q}) =
\hbar{\sqrt{4{C(Q_{max})}/m}}|\sin({Q}{\sigma}/2)|
\eqno (49)$$

\noindent
It is evident that phononlike dispersion is expected till
the excitation wavelength ${\Lambda}$ remain larger than
$d$ ({\it i.e.} $Q < 2\pi/d$).  However, the momentum and
energy of the excitation would be carried by only a single
particle, if $Q < 2\pi/d$ ({\it i.e.} ${\Lambda} < \sigma$)
and this implies that $E(Q)$ for $Q > 2\pi/d$ would follow  
$${E_{sp}(Q)} = {\hbar^2{Q}^2/2{m_F}}  \eqno (50)$$ 

\noindent
which represents a kind of {\it single particle} dispersion
with $m_F$ being a kind of {\it effective mass that measures
the effect of quantum correlation in} ({\bf q}, -{\bf q})
pair configuration.  It is expected to be around $4m$ near
$Q = 2\pi/d$ and then decrease slowly with increasing $Q$
beyond this point.  
  
\bigskip
The transition of $E(Q)$ from ${E_{ph}(Q)}$ to ${E_{sp}(Q)}$
would, obviously, occur at $2\pi/d < {Q} < 2\pi/{\sigma}$,
before the ${E_{ph}(Q)}$ (Eqn. 49) meets its zero value at  
$2\pi/\sigma$.  This implies that $E(Q)$ has to have its
minimum ({\it identified as the roton minimum} for $He-II$)
at a $Q = {Q_{min}}$ near the mid-point of 
$Q = 2\pi/d$ and $Q = 2\pi/{\sigma}$, {\it i.e.}  
$${Q_{min}} \approx  \left[\frac{\pi}{d} + 
\frac{\pi}{\sigma}\right] \eqno (51)$$
 
\bigskip
Evidently, Eqns. 48-51 represent a Landau type spectrum
and use of Eqn. 47 in Eqn. 46 renders  
$$v_p = v_g = {\sqrt\pi}h/2md   \eqno (52)$$ 

\noindent   
where $v_p$ and $v_g$, respectively, represent the phase
and group velocities of phonons at $Q \approx 0$.

\bigskip 
We further find that the equation of motion of $r_s$ (the
$r$ of $s$-th atom)
$${{\partial}^2_t}r_s  =  - {\frac{1}{4}}{\omega}_o^2
[2r_s - r_{s-1} - r_{s+1}]    \eqno (53)$$ 
transforms into a similar equation for $p_s = \hbar{q_s}$ by  
operating $m{\partial}_t$ and this concludes
$${\omega}_r({Q}) = {\omega}_q({Q}) \eqno (54)$$

\noindent
implying that the omon dispersion ${\omega}_q({Q})$ is not
different from phonon dispersion ${\omega}_r({Q})$.
Interestingly, as concluded in Section (5.0), an omon is an
anti-phonon quantum quasi-particle.  

\bigskip

\centerline{\bf 5.0  Energy Gap, Self Energy and Bound Pairs}

\bigskip
The system at $T \le T_{\lambda}$ is separated into
two components, F1 and F2 ({\it cf.} Section 2.6).
While F1 depicts $K-$motions as quasi-particle excitations
whose density decreases on cooling and vanishes at $T=0$,
F2 represents the G-state of the system and portrays
zero-point $q-$motions with $q = q_o$.  Evidently, the
unique low $T$ properties of our system that last even
at $T=0$, are related to F2 and to this effect we examine
its evolution on cooling the system from $T_{\lambda}$
to $T=0$.

\bigskip
As concluded in Sections 2.5 and 2.6, particles in the
system are locked at $\varepsilon_o(T_{\lambda})$.
However, when the system is cooled to $T \le T_{\lambda}$,
$\varepsilon_o = h^2/8md^2$ tends to have a value lower
than $\varepsilon_o(T_{\lambda})$ which, obviously,
requires an increase $d$ from $d_{\lambda}$ to
$d_T = d_{\lambda} + \Delta d$; here $d_{\lambda}$ means
$d$ at $\lambda-$point.  Naturally, this brings zero-point
force ($f_o = - \partial_d \varepsilon_o = h^2/4md^3$) into
operation which increases $d$ by pushing two neighboring
particles to an increased distance $d_T$ against their '
inherent attractive force $f_a$.  Consequently, the system
has volume expansion on cooling around $T_{\lambda}$ and
this agrees exactly with experimentally observed negative
thermal expansion co-efficient of liquid $^4He$ at
$T \le T_{\lambda}$ [1].  Evidently, the evolution of F2
on cooling is an inter-play of $f_o$ and $f_a$.  In the
following we use this inference and other important aspects
of F2 to present three inter-related pictures to conclude
the formation of ({\bf q}, - {\bf q}) bound pairs, energy
gap and self energy.

\bigskip
\noindent
{\bf (1)}:  The impact of $f_o$ on the states of the
system at $T \le T_{\lambda}$, which is decided by its
balance with $f_a$, can be understood by using $V(r_{ij})$
as a perturbation.  We first demonstrate this for the
states of a pair of particles and to this effect
diagonalise (2x2) energy matrix defined by $E_{11} =
E_{22} = \varepsilon_o$ and $E_{12} = E_{21} = \beta_o$
with $\beta_o = <V(r_{ij})>$.  We note that $f_o$ is a
consequence of $V_{HC}(r)$ clubbed with wave nature of
particles which increases its operation from $\sigma$
to $\lambda/2$, particularly, for the particles of
$\lambda/2 > \sigma$ [59].  Naturally, it appears on the
scene only at $T \le T_{\lambda}$ where
${\lambda}/2 \to d^+_{\lambda}$ (little more than $d$
decided by $V(r_{ij})$.  The macro-orbitals of
particles under these physical situations should,
obviously, have their overlap to render non-zero
$<V(r_{ij})>$ and this is in line with our zero-order
results $<V_{HC}(r)> = 0$.  Evidently, one finds that the
pair has two states of energy $\varepsilon_o
\pm |\beta_o|$; in fact $|\beta_o|$ should better be
replaced by $|\beta_o(T)|$ as the overlap  of the
macro-orbitals of two particles may depend on $T$.  The
states of energy $(\varepsilon_o - |\beta_o(T)|)$ and
$(\varepsilon_o + |\beta_o(T)|)$ can, respectively, be
identified as bonding (or paired) and antibonding (or
unpaired) states.  This follows the established approach of
{\it Molecular Orbital Theory} [75] applied to a similar
case in which two identical atomic orbitals form two
molecular orbitals of bonding and anti-bonding nature.  The
pair is expected to be in bonding state provided the two
particles remain locked in the relative configuration
characterized by Eqn. 14.

\bigskip
Now we apply the same approach to the state of $N$ particles
by constructing a $N$x$N$ matrix for $H_r(N)$ (representing
the relative motion of $N$ particles) with $[H_r(N)]_{mn} =
\varepsilon_o$ for $m=n$, and $[H_r(N)]_{mn} =
[V(r)]_{mn}$ for $m \not = n$ with $[V(r)]_{mn}$
having non-zero value only if $m$ and $n$ refer to two
neighboring particles.  Note that each particle has 6-12
(depending on the symmetry of their assumed spatial
arrangement) nearest neighbors. The diagonalisation of this
matrix renders N/2 energy levels with energy $>$
$N\varepsilon_o{(T_{\lambda})}$ ({\it the anti-bonding
states}) and N/2 energy levels with energy $<$
$N\varepsilon_o{(T_{\lambda})}$ ({\it the bonding states}).
It is natural that F2, separated from F1, assumes the
lowest possible bonding state.  Since the perturbative
effect of $V(r)$ lowers $\varepsilon_o$ of each particle,
identically, all particle fall in the bonding state
simultaneously and acquire a kind of collective binding.

\bigskip
\noindent
{\bf (2).} Although, the preceding analysis renders a good
account for the origin of the collective binding of
particles, however, a relation that helps in calculating its
magnitude is needed.  To this effect, we note that the
energy of F2 deceases from $N\varepsilon_o(T_{\lambda})$ to
$N.\varepsilon_o{(T)}$ when the system moves from its
${\lambda}-$point to $T < T_{\lambda}$.  This concludes that
energy of each particle falls by
$$\Delta\epsilon = [\varepsilon_o(T_{\lambda}) -
\varepsilon_o(T)] \approx \frac{h^2}{4md^2_{\lambda}}
\frac{d_T - d_{\lambda}}{d_{\lambda}}
= 2\varepsilon_o\frac{d_T - d_{\lambda}}{d_{\lambda}}
\eqno (55) $$

\noindent
A simple analysis of the equilibrium between $f_o$ and $f_a$
reveals that half of the $\Delta\epsilon$ is stored in
the system as strain energy in inter-particle bonds, while
the remaining half is lost to surrounding.  Since this happens
to all particles, the net energy loss ($ = N\Delta\epsilon/2$)
by F2 turns out to be their collective binding.  Using
the coherence property of the system (evident from
$\Delta\phi = 2n\pi$), one may  find that the effective
binding per particle becomes much larger than $K_BT$ [76],
even when the real binding per particle is very small and
this ensures the stability of our system with collective
binding among its particles.  Evidently, the entire system
at $T < T_{\lambda}$ seems to represent a macroscopically
large single molecule as envisaged by Foot and Steane [77]
for the BEC state of trapped dilute gases.  It is obvious
that the stability of this state can not be disturbed by
any perturbation of energy $< N\Delta\epsilon/2$ and we may
define
$$E_g(T) = N\Delta\epsilon/2
= N\varepsilon_o\frac{d_T - d_{\lambda}}{d_{\lambda}} 
\eqno (56)$$

\noindent
as an energy gap between the S- and N-states.

\bigskip
The strain energy $\Delta V_s = N\Delta\epsilon/2$, stored
with the elongated bonds between nearest neighbors,
obviously, represents the net increase in the potential
energy (= $\Delta V_s$) of particles and we call it as the
self energy of the system.  Since the strain
($\Delta d/d_{\lambda} = (d_T - d_{\lambda})/d_{\lambda}$
arising due to increase in quantum size
of each particle from $d_{\lambda}$ to $d_T$) is an obvious
function of the momentum $q$ of each particle, we have
$\Delta V_s = \Delta V_s (q_1, q_2, ... q_N)$ and this
prepares the system to sustain phononlike waves of
collective oscillations of momentum coordinates of particles
(named as omons, {\it cf.} Section-4.2).  Evidently, 
$\Delta V_s$ can also be recognized as the energy of omon
field,  The fact that $\Delta V_s$ increases with decreasing
$T$ implies that omon field intensity increases when phonon
field intensity decreases and {\it vice versa} and this
means that either the omon field intensity is the intensity
of phonons condensed with the system or an omon is an
{\it anti-phonon} quantum quasi-particle.
We note that $\Delta V_s$ assumes it maximum value at
$T = 0$ and serves as the source of collective motions at
$T = 0$ when phonon cease to exist.   

\bigskip
Since each particle in our system represents a
({\bf q}, - {\bf q}) pair, the above analysis concludes
that all particles at $T \le T_{\lambda}$ assume the
states of ({\bf q}, - {\bf q}) bound pairs.  Further in
view of our results ({\it cf.}, Eqn. 14), which indicate
that each pair of particles in F2 is locked at $<k> = 0$,
$<r> = d$ and $<\phi> = 2n\pi$, our {\it bound pair}
represents two particles bound in all the three ($k-$, $r-$
and $\phi$) spaces.  However, this does not imply that two
particles in the system form a $He_2$ type diatomic
molecule; the entire system assumes a state where particles
interact with their neighbors identically by two body
forces.  We use the word bound pair because each particle
is in a state of ({\bf q},-{\bf q}) pair.

\bigskip
\noindent
{\bf (3)}. As inferred by Ulhenback and Gropper [66],
a quantum system can be treated like a classical system
by adding quantum correlation potential to the hamiltonian
of the system.  Using this inference and Eqn. 16, we find
an alternative relation for $E_g(T)$.  We note that F2, for
its proximity with F1, has
a small number of particles $N^*(T)$ in thermally excited
$q-$motion states of $q \ge 2q_o$.  These particles are,
obviously, devoid of quantum correlations with particles in
the G-state ($q = q_o$).  However, in the process of
cooling, these particles fall to the G-state and
establish quantum correlations with other particles of this
state.  Consequently, the energy of each particle falls
below zero energy level by $-k_BT_o\ln{2}$.  Assuming that
number of these particles decreases from $N^*(T_{\lambda})$
to $N^*(T)$, we find
$$\Delta{\epsilon}(T) =
- k_BT_o\ln{2}[N^*(T_{\lambda}) - N^*(T)] \eqno (57)$$ 

\noindent
as the net fall in energy due to quantum correlations.  Here
we have
$$N^*(T) = {V\over 4\pi^2}
\left[\;{2m\over\hbar^2}\;\right]^{3/2}
\int\limits^{\infty}_{\varepsilon_c}
\left[\;\exp{\left(\frac{\varepsilon
- \varepsilon_o}{k_BT}\right)} - 1\;\right]^{-1}
{\sqrt \varepsilon}d{\varepsilon} \eqno (58)$$

\noindent
where $\varepsilon_{c} = \hbar^2Q^2_c/2m$ (with $Q_c \approx
2\pi/\sigma$).  Once again it is evident that half of the
$\Delta{\epsilon}(T)$ energy released out is stored back
in the system as strain energy in expanding nearest
neighbor inter-particle bonds which indicates that the
net fall in energy per particle is
$\frac{1}{2}\Delta{\epsilon}(T)$.  Since the negative
value of $\frac{1}{2}\Delta{\epsilon}(T)$ means a kind
of collective binding of particles and we have $E_g(T)
= \frac{1}{2}\Delta{\epsilon}(T)$ is an alternative
relation for $E_g(T)$.  Its accuracy is corroborated by
the fact that $\frac{1}{2}\Delta\epsilon{(T)}$ (Eqn. 57) and
$E_g(T)$ (Eqn. 56) have closely equal value for $He-II$ at
all $T < T_{\lambda}$ [56].  Note that Eqn. 58 is an
approximate relation for $N^*(T)$, since it uses a free
particle dispersion, $\varepsilon = \hbar^2Q^2/2m$ which
is valid only to a good approximation.

\bigskip
\centerline{\bf 6.0  Energy Gap and its Consequences }

\bigskip
In what follows from the Section-5.0, the energy of F2 at
$T \le T_{\lambda}$ can be expressed as $F(q) =
N\varepsilon_o(T_{\lambda}) - E_g(T)$ where
$N\varepsilon_o(T_{\lambda})$ is constant.  Evidently, the
origin of different properties (including superfluidity and
related aspects) of our system lies with $E_g(T)$ and this
fact is used for the following analysis.
 
\bigskip
\noindent
{\it 1. Superfluidity and Related Properties} :   

\bigskip
If two heads $X$ and $Y$ in the system have small $T$ and
$P$ (pressure) differences, the equation of state is
$E_g(X) = E_g(Y) + {\rm S}{\Delta}T - {\rm V}{\Delta}P$.
Using $E_g(X) = E_g(Y)$ for equilibrium, we get
$${\rm S}{\Delta}T = {\rm V}{\Delta}P     \eqno (59)$$

\noindent
This reveals that : (i) the system should exhibit
thermo-mechanical and mechano-caloric effects, and (ii) the
measurement of $\eta$ by capillary flow method performed
under the condition ${\Delta}T = 0$ and of thermal
conductivity $(\Theta)$ determined under ${\Delta}P = 0$
should reveal $\eta = 0$ and $\Theta \approx \infty$,
respectively.  As such the S-phase is expected to be a
superfluid of infinitely high $\Theta$.
        
\bigskip
Interestingly, several important aspects of our system can
also be followed qualitatively from the configuration of F2.
For example, we note that: (i) a close packed arrangement of
particles in a fluid like system can have no vacant site,
particularly, because two neighboring particles experience
zero point repulsion which, naturally, means that the
system should have very large $\Theta$, (ii) the system can
not have thermal convection currents for its large $\Theta$
and close packing of particles and this explains why $He-II$
does not boil like $He-I$, (iii) since particles in F2 can
move only in order of their locations and cease to have
relative motion, the system is bound to exhibit vanishingly
small $\eta$, particularly, for their flow in narrow
capillary, {\it etc.})   In the rotating fluid, however,
particles moving on the neighboring concentric circular
paths of quantized vortices have relative velocity as a
source of natural viscous behavior. This explains both
{\it viscosity and rotation paradoxes} [5].  As such
the loss of viscosity in linear motion is not due to
any loss of viscous forces among the particles, rather
it is the property of the S-phase configuration in which
particles cease to have relative or collision motion. 

\bigskip
\noindent
{\it 2. Critical Velocities and Stability of S-phase} :  

\bigskip
Using the same argument, which renders Eqn. 44, we find
that the state function $\Phi_o{(N)}$ of the S-phase
changes to $\Phi_o^*{(N)}$ when the system is made to
flow with velocity $v_f = \hbar\Delta{\bf q}/m$.  We have
$$\Phi_o^*{(N)} = \Phi_o{(N)}\exp{(i{\bf K}.
\Sigma_i^N{\bf R_i})}\exp{[-i[N(\varepsilon_o +
\varepsilon{(K)})-E_g(T)]t/\hbar]}
\eqno (60) $$ 

\noindent
with $2\Delta{\bf q} = {\bf K}$. This reveals that the
S-state function remains stable against the flow unless
its energy $Nmv_f^2/2 = N\varepsilon{(K)}$ overtakes
the collective binding $E_g(T)$.  We use this observation
to explain critical velocity $v_c$ for which the system
loses superfluidity.   Equating $E_g(T)$ and flow energy
($N.mv_f^2/2$ with $v_f = v_c$), we obtain the upper
bound of $v_c$.  We have 
$$v_c(T) = \sqrt{[2E_g(T)/Nm]} \eqno (61) $$
\noindent 
A $v_c < v_c(T)$, at which the superfluid may show signs of
viscous behavior, can be expected due to creation of
quantized vortices.  However, this cause would not destroy
superfluidity unless energy of all vortices produced in the
system exceeds $E_g(T)$.

\bigskip
\noindent
{\it 3. Coherence Length} :

\bigskip
Since the main factor responsible for the coherence of F2
is its configuration which locks the particles at
$\Delta\phi = 2n\pi$ ({\it cf.} Eqn. 14) with collective
binding $E_g(T)$, the {\it coherence length} (not to be
confused with healing length [5]), can be obtained from
$$\xi{(T)} = 1/mv_c(T)
= h{\sqrt{[N/2mE_g(T)]}} \eqno (62). $$

\bigskip
\noindent
{\it 4. Superfluid Density} :   

\bigskip
Correlating the {\it superfluid density}, ${\rho}_s$, as the
order parameter of the transition, with $E_g(T)$ we find a 
new relation  
$${\rho}_s(T) = \frac{E_g(T)}{E_g(0)}\rho{(T)} =
\frac{d_T - d_{\lambda}}{d_o - d_{\lambda}}\rho{(T)}
\eqno (63) $$

\noindent
to determine  
${\rho}_s{(T)} $ and {\it normal density}, ${\rho}_n{(T)}
= {\rho}{(T)}  - {\rho}_s{(T)}$. Evidently, $v_c(T)$,
$\xi{(T)}$, and ${\rho}_s{(T)}$ can be obtained if we know
$E_g(T)$ (Eqn. 56).  Further since S-state function vanishes
at the boundaries of the system, it is natural that $E_g(T)$
and ${\rho}_s{(T)}$ also vanish there.

\bigskip
\noindent
{\it 5. Superfluid Velocity} :

\bigskip
Concentrating only on the time independent part, Eqn. 60
can be arranged as
$$\Phi_o^*{(N)} = \Phi_o{(N)}\exp{(iS(R)}   \eqno (64) $$ 

\noindent
with 
$$S(R) = {\bf K}.(\Sigma_i^N{\bf R}_i)       \eqno (65)$$

\noindent
being the phase of the S-state.  This renders   
$${\bf v}_s =  \frac{\hbar}{2m}{\bigtriangledown}_{R_j}S(R)
= \frac{\hbar\Delta{\bf q}}{m}
\eqno (66)$$    

\noindent
as a relation for the superfluid velocity; here we use the
fact that ${\bigtriangledown}_{R_j}S(R)$ renders the
momentum of the pair (not of a single particle).  One may
find that Eqn. 66 does not differ from the superfluid wave
function presumed in the $\Psi-$theory of superfluidity [78]
({\it cf.} Section (2.3) of Reference [8]), of course for
the well defined phenomenological reasons, $\Psi-$theory
assumes $S(R)$ to be a complex quantity.

\bigskip
\noindent
{\it 6. Quantized Vortices}

\bigskip
Using the symmetry property of a state of bosonic system,
Feynman [54, 71] showed that the circulation, $\kappa$, of the
velocity field should be quantized, ${\it i.e.}$, $\kappa
= \frac{nh}{m}$ with $n = 1,2,3,...$  However, Wilks [1]
has rightly pointed out that this account does not explain
the fact that $He-I$ to which Feynman's argument applies
equally well, does not exhibit quantized vortices.  Using
Eqn. 66, we find that
$$\kappa = \sum_i {\bf v}_s(i).\Delta{\bf r}_i =
\frac{\hbar}{m}\sum_i \Delta{\bf q}_i.\Delta{\bf r}_i
= \frac{nh}{m} \eqno (67) $$

\noindent
by using the condition that $\sum_i \Delta{\bf q}_i
.\Delta{\bf r}_i = 2n\pi$ which presumes that particles
moving on a closed path maintain phase correlation.
To this effect our theory reveals that particles have their
$\phi$-positions locked at $\Delta{\phi} = 2n\pi$ only in
S-phase which indicates that only this phase can exhibit
quantized vortices.  However, since particles of N-phase
have random distribution ($\Delta{\phi} \ge 2n\pi$) in
$\phi$-space, this phase can not sustain $\phi-$correlation
and quantized vortices.

\bigskip
\noindent
{\it 7. S-state and its Similarity with Lasers}

\bigskip
We note that the system below $T_{\lambda}$ defines a 3-D
network of SMWs extending from its one end to another
without any discontinuity.  In lasers too these are the
standing waves of electromagnetic field that modulate the
probability of finding a photon at a chosen phase point.
The basic difference between the two lies in the number of
bosons in a single {\it anti-nodal region} (AR) of a SMW.
In case of lasers this could be any number since photons
are non-interacting particles but for a SIB like $^4He$
or $^{87}Rb$ one AR can have only one atom.

\bigskip
\centerline{\bf 7. Consistency with Phenomenological
Theories and Experiments}

\bigskip
\noindent
{\it 1.  Two fluid theory}

\bigskip
Since our system at $T \le T_{\lambda}$ has two separated
components F1 and F2, it can be identified to be a
homogeneous mixture of two fluids.  While F1, described by
plane waves of momentum $K$, represents a gas of
non-interacting quasi-particle excitations, F2 representing
zero-point $q-$motions describes the G-state of the
system where particles are locked with $<k> = 0$, $<r> =
\lambda/2$ and $\Delta\phi = 2n\pi$ and consequently cease
to have relative motion, or collisional motion essential
for non-zero viscosity.  Naturally, the thermodynamic
properties, such as specific heat, non-zero entropy,
{\it etc.}, as well as non-zero viscosity are contributed
totally by F1.  As such F1 has all properties of a N-fluid,
while F2 having zero entropy, zero viscosity, {\it etc.} has
the properties of S-fluid.  Most interestingly our theory
provides microscopic origin to the two fluid theory of
Landau [14] and in spite of its separation into F1 and F2,
each and every particle can be seen to participate in both
fluids.

\bigskip
\noindent
{\it 2.  $\Psi-$ Theory }

\bigskip
We find that superfluidity is basically a property of the
G-state (F2 component) of our system represented by
Eqn. 13 which can also be expressed as
$$\Phi_o(N) = \sqrt{n}  \eqno (68) $$

\noindent
with $n = N/{\rm V}$.  However, when F2 is made to flow its
state is given by $\Phi^*_o(N) = \Phi_o(N)\exp{iS(R)}$
(Eqn. 64).  For the phenomenological reasons (such as the
superfluid density $n_s$ is not always equal to $n$), we
assume that the phase $S(R)$ (Eqn. 65) is a complex quatity
given by
$$S(R) = \xi_r(R) + i\xi_i(R)   \eqno (69)$$ 

\noindent
which renders $n_s = n\exp{(-2\xi_i(R)})$.  Evidently,
$\Phi^*_o(N)$ has the structure of $\Psi-$function that
forms the basis of the well known $\Psi-$theory of
superfluidity.  This shows that our theory provides
microscopic foundation to the highly successful
$\Psi-$theory [78].

\bigskip
\noindent
{\it 3. Properties of He-II}

\bigskip
\noindent
(i) {\it Thermodynamic properties :} An estimate of
$T_{\lambda}$ for LHE-4 by using Eqn. 31 (with $m^* = m$)
renders a value falling around 2.1 K which agrees closely
with experimental value 2.17 K.  In view of an
observation by Woods and Cowley [4], Feynman's relation
$E(Q)_{Feyn} = \hbar^2Q^2/2mS(Q)$ [71] renders $E(Q)$
values that are nearly two times of $E(Q)_{expt}$
(experimental $E(Q)$ for $He-II$).  Evidently, our $E(Q)$
(Eqn. 45) equaling one half of the $E(Q)_{Feyn}$, ensures
good matching with $E(Q)_{expt})$ for He-II and this
implies that our theory can explain the thermodynamic
properties of $He-II$ accurately.  In addition, using our
alternate set of relations for $E(Q)$ (Eqns.46-51),
concluded by our theory, we find good agreement between
theory and experiment for : (i) for the anomalous nature
of phonon dispersion at low $Q$, velocity of sound at
low $Q$ [56 and 80].

\bigskip
The problem of explaining the experimentally observed
logarithmic singularity in $C_p(T)$ of LHE-4 at $T_{\lambda}$
has been a challenging task ({\it cf.} a remark by Feynman in
his book [71, p.34]).  Naturally, the fact that for the first
time our theory succeeded in explaining it at quantitative
level speeks for the accuracy of the microscopic foundation 
of our theory.  Using the parameters of liquid ${^4He}$
in Eqn. 42 with $\nu = 0.55$ and $\delta{\phi}_{\lambda}(o)
= \pi$, we find
$$C_p({\rm J/mole.K}) \approx
 -5.71\ln|\zeta| - 10.35 = -A.\ln|\zeta| + B  \eqno (70)$$

\noindent
which can be compared with experimental results
$C_p = -5.355\ln|\zeta| - 7.77$ for $T > T_{\lambda}$ and
$C_p = -5.1\ln|\zeta| + 15.52$ for $T < T_{\lambda}$ [79].
The fact that our A value agrees closely with experiments
speaks of the accuracy of our theoretical result.  With
respect to our choice of $\nu = 0.55$ and
$\delta{\phi}_{\lambda}(o) = \pi$, we note that:
(i) $\delta{\phi}_{\lambda}$ originates basically
from change in momentum ${\Delta}k \approx {\xi}^{-1}$ and
$\xi$ varies around $T_{\lambda}$ as
$|T - T_{\lambda}|^{-\nu}$; note that critical exponent
$\nu$ lies in the range 0.55 to 0.7 [70].  For our choice
of $\delta{\phi}_{\lambda}(o) = \pi$, we note that
$\pi$ is the largest possible value by which phase
position of a particle changes.  We note that experimental
value of $B = - 7.77$ (for $T > T_{\lambda}$) and
+ 15.52 (for $T < T_{\lambda}$) can be obtained if 
$\delta{\phi}_{\lambda}(o)$ is chosen to be around
$0.78\pi$ and $0.08\pi$, respectively. These are
consistent with the reasons of assymetry of this
singularity as stated in Section-3.6 and the fact that
$\phi-$fluctuations at $T > T_{\lambda}$ are expected
to be much higher than those at $T < T_{\lambda}$.

\bigskip
\noindent
(i) {\it Hydrodynamic properties :} Our theory provides
microscopic foundations for: (i) two fluid theory
of Landau (Section 7.1), (ii) the $\Psi-$theory of
superfluidity (Section 7.2), (iii) the quantized
vortices (Section 6.6), (iv) the vanishing of $\rho_s(T)$
at the boundaries of the system (Section 6.4), (v) the
collective binding or energy gap (Section 5.0) which not
only ensures the stability of S-state but also
accounts for the critical velocities, 
coherence length, {\it etc.} (Section 6.0).  Evidently, it
can explain the hydrodynamic properties of superfluid SIB 
like $He-II$.  While a brief analysis of the agreement
between theory and experiments for $He-II$ is available in
[56], a detailed quantitative study, being completed by
our group [80], would be published soon.

\bigskip
As shown Fig.(1), we note that the $T$ dependence of: (i)
the fraction of particles condensed in $K = 0$ state,
$n_{K=0}(T^*)$ [Curve A$^*$, Eqn. 37] and (ii) $n^*(T) =
[1 - N^*(T)/N^*(T_{\lambda})]$ [Curve B, Eqn. 58] have an
excellent agreement with similar nature of the order
parameter ($\Omega{(T^*)}$ [Curve C$^*$, Eqn. 36] of a
second order transition.  In addition they have similar
agreement with the $T$ variation of : (i) $\rho_s{(T)}$
(Curve E2) deduced from the experimentally observed second
sound velocity of $He-II$ (taken from [5]), and (ii) the
normalized inter-particle bond strain $d_T - d_{\lambda}/
d_o - d_{\lambda}$ (curve E1) obtained from experimental
values of $\rho{(T)}$ [1].  These facts, evidently, identify
inter-particle bond strain as a basic aspect of the
$\lambda-$transition, $T^*$ as the effective temperature scale
for the order parameter, and establish the accuracy of our
theory in accounting for the unique properties of $He-II$.
It is further corroborated by the fact that E1 and E2
curves do not agree with $n_{K=0}(T) = 1-
(T/T_{\lambda})^{3/2}$ (Curve A) and $\Omega{(T)}$ [Curve C,
Eqn. 35].  As such an overall agreement between
experimental and theoretical results for LHE-4 establishes
the accuracy of our theory.
 
\bigskip
\centerline{\bf 8.0  Concluding Remarks}

\bigskip
This paper presents an alternative approach to the
microscopic understanding of a SIB.  We find that each
particle in the system has two motions, $q-$ and $K-$,
because it represents a pair of particles moving with equal
and opposite momenta ({\bf q}, -{\bf q}) at their CM which
moves with momentum {\bf K}.  Consequently, it should be
described by a {\it macro-orbital}. The onset of $\lambda-$
transition is an order-disorder of particles in $\phi-$space
followed simultaneously by their BEC in the state of $q =
q_o$ and $K = 0$.  It is a consequence of quantum correlations
which drive $q-$values towards $q = q_o$ and $K-$values
towards $K = 0$ and an inter-play of zero-point force
$f_o$ and inter-particle attraction $f_a$ which leaves a
kind of mechanical strain
in the inter-particle bonds as the basic component of order
parameter (Cf. Fig.1 curve E1). The system is expected
to exhibit $-ve$ thermal expansion coefficient around
$T_{\lambda}$ as evinced by LHE-4.

\bigskip
The transition separates the two motions and the system
behaves like a homogeneous mixture of two fluids: (i) F1,
-representing a gas of quasi-particle excitations
describing the plane wave $K-$motions of $N$ particles and
(ii) F2, -representing the G-state of the system where
particles are locked at $<r> = d$, $\Delta\phi= 2n\pi$ and
zero-point motions at $q = q_o$ and $K = 0$.  The condensate
fraction $n_{K=0}(T)$ rises smoothly from $n_{K=0}(T) = 0$
at $T = T_{\lambda}$, reaches $n_{K=0}(T) \approx 1.0$
around $T = T_o$ and $n_{K=0}(T) = 1.0$ at $T = 0$.  The HC
interaction leading to excluded volume condition [51]
pushes all particles to occupy identically equal volume and
a state of identically equal $q = q_o$. 

\bigskip
Superfluidity and related properties are, basically,
the properties of $T=0$ state of F2 component of a SIB
which implies that $\lambda-$transition is a kind of quantum
transition which occurs at non-zero $T$ for the proximity
of F2 with F1.  Particles in F2 represent ({\bf q}, -{\bf q})
bound pairs, not only in $q-$space but also in $r-$ and
$\phi-$spaces.  They assume a kind of close packed
arrangement with collective binding which leads the entire
system to behave like a macroscopically large single
molecule.  The collective binding represents an energy gap
between the S- and N-fluid phases.  The
S-state is consistent with microscopic uncertainty
as evident from
$q \ge \pi/d$ as well as macroscopic uncertainty since
$\Phi_o(N)$ (Eqn. 13) vanishes at the boundaries of the
system.  The difference between the S- and N-fluid
states can be identified with a difference in the
ordered positions and motions of soldiers in an organized
army platoon and a crowed.  Our theory makes no assumption about
the nature of BEC in a SIB.  It reaches its all conclusions
from the analysis of the solutions of its $N-$body
Schr\"{o}dinger equations.

\bigskip
In many respect, above summarized inferences of our theory
differ from conventional theories which presume the existence
of $p=0$ condensate in the S-state of a SIB.  Using
two different theoretical approaches ({\bf A1} and {\bf A2}
summarized in Section-1.0) and different possible mathematical
tools, they conclude a maximum of $\approx 13\%$ $p=0$ condensate
in $He-II$ and $\approx 60\%$ in recently discovered BEC state
of dilute trapped gases [81] and for the last several decades
this conclusion has been regarded as basic origin of the unique
properties of every superfluid SIB.

\bigskip
In summary, this paper provides an alternative theoretical
framework concluding a different picture of S-phase
of a SIB like LHE-4.   Analyzing the wealth of experimental
observations on the S-state of different SIB(s),
one may agree that the wave nature of particles has wonderful
capacity for their self organization in phase space at
$\phi = 2n\pi$ with $<r> = \lambda/2 =d$ and our theoretical
framework can reveal the truth of the low $T$ behavior
of widely different many body systems.

\bigskip
Finally, as discussed briefly in [57, 60], the framework of
our theory unifies the physics of widely different many
body systems of interacting bosons and fermions including
low dimensional systems, atomic nucleus, newly discovered
BEC states [81], {\it etc.}  This is corroborated by our detailed
studies of $N$ HC particles in 1-D box [61] and microscopic
theory of superconductivity [62] which not only accounts for
the highest $T_c$ that we know to-day but also reveals that
the phenomenon may, in principle, occur at room temperature
and co-exist with ferro-magnetism.  For the first time our
theory concludes [60, 62] that superfluid transition in
liquid $^3He$ should occur around 2mK which matches almost
exactly with experimental result.  It may be emphasized
that ({\bf q}, -{\bf q}) bound pairs of particles as
concluded in [62] and the present paper should not be confused
with Cooper pairs [82] since the two have several similarities
and dissimilarities.

\newpage

\centerline{\bf Appendix -A}

\bigskip
\centerline{\bf Wave Mechanics of Two HC Particles and Macro-orbitals}

\centerline{{\it detailed analysis on this topic avaible in} [62]}

\bigskip
The dynamics of two HC impenetrable particles P1 and P2 can
be described by
$$\left(-\frac{\hbar^2}{2m}\sum_i^2\bigtriangledown^2_i +
V_{HC}(r)\right)\psi{(1,2)} = E(2)\psi{(1,2)} \eqno(A-1)$$
 
\bigskip
As argued in Section 2.1, we may use $V_{HC}(r) \equiv
A\delta{(r)}$ (Eqn. 3) and simplify the solution by using
CM coordinates
$${\bf r}= {\bf b}_2 - {\bf b}_1 \quad {\rm and} \quad
{\bf k} = {\bf p}_2 - {\bf p}_1 = 2{\bf q}, \eqno(A-2)$$

\noindent
where {\bf r} and {\bf k}, respectively, represent the
relative position and relative momentum of P1 and P2,
and
$${\bf R}= ({\bf b}_2 + {\bf b}_1)/2  \quad {\rm and} \quad
{\bf K} = {\bf p}_2 + {\bf p}_1, \eqno(A-3)$$

\noindent
where {\bf R} and {\bf K}, similarly, refer to the
position and momentum of their CM.  Without loss of
generality, Eqns. A-2 and A-3 also render
$${\bf p}_1 = - {\bf q} + \frac{\bf K}{2} \quad {\rm and}
\quad {\bf p}_2 = {\bf q} + \frac{\bf K}{2}.
\eqno(A-4)$$
 
\noindent
We use these relations to express Eqn. A-1 as     
$$\left(-\frac{\hbar^2}{4m}\bigtriangledown^2_R
-\frac{\hbar^2}{m}\bigtriangledown^2_r +
A\delta{(r)}\right)\Psi{(r,R)} = E(2)\Psi{(r,R)} \eqno(A-5)$$

\noindent
with 
$$\Psi{(r,R)} = \psi_k(r)\exp(i{\bf K}.{\bf R}). \eqno(A-6)$$

\noindent
The HC interaction affects only $\psi_k(r)$ (the state
of relative motion) which represents a solution of
$$\left(-\frac{\hbar^2}{m}\bigtriangledown^2_r +
A\delta{(r)}\right)\psi_k(r) = E_k\psi_k(r) \eqno(A-7)$$

\noindent
with $E_k = E(2)-\hbar^2K^2/4m$, while the CM motion
$[\exp(i{\bf K}.{\bf R})]$ remains unaffected. 

\bigskip
To find $\Psi{(r,R)}$ (Eqn. A-6), we treat
$A\delta{(r)}$ as a step potential.  Since $A\delta{(r)}
= 0$ for $r \not = 0$, P1 and P2 can be represented,
by independent plane waves but at $r = 0$ where
$A\delta{(r)} = \infty$, we use the condition
$\Psi{(r,R)}|_{r=0} = 0$.  Following these points,
a state of P1 and P2 can, in principle, be expressed by  
$$\Psi{(1,2)}^{\pm} =
\frac{1}{\sqrt{2}}[u_{{\bf p}_1}({\bf r}_1)u_{{\bf p}_2}({\bf r}_2)
\pm u_{{\bf p}_2}({\bf r}_1)u_{{\bf p}_1}({\bf r}_2]. \eqno(A-8)$$

\noindent
However, $\Psi{(1,2)}^+ $ (of $+ve$ symmetry for
the exchange of two particles) has to be excluded as
undesirable function because it does not satisfy
$\Psi{(r,R)}|_{r=0} = 0$.  But this leaves only
$\Psi{(1,2)}^-$ of $-ve$ symmetry which does not fit
with the bosonic character of our system.  We encountered
this problem in our recent study of the 1-D analogue of
Eqn. A-7 in relation to the wave mechanics of two HC
impenetrable particles in 1-D box [59].  Following this
study we find that the state of P1 and P2 can be expressed
by
$$\zeta{(r,R)}^{\pm} =
\zeta_k(r)^{\pm}\exp{(i{\bf K}.{\bf R})}  \eqno(A-9)$$

\noindent
with 
$$\zeta_k(r)^- = \sqrt{2}\sin{({\bf k}.{\bf r}/2)} \eqno(A-10)$$ 

\noindent
of -ve symmetry, and 

$$\zeta_k(r)^+ = \sqrt{2}\sin{(|{\bf k}.{\bf r}|/2)} \eqno(A-11)$$

\noindent
of +ve symmetry.  

\bigskip
$\zeta_k(r)^{\pm}$ represents a kind of {\it stationary
matter wave} (SMW) which modulates the probability
$|\zeta_k(r)^{\pm}|^2$ of finding two particles at their
relative phase ($\phi$) position $\phi = {\bf k}.{\bf r}$
in the $\phi-$space.  Interestingly, the equality
$|\zeta_k(r)^-|^2 = |\zeta_k(r)^+|^2$ renders an
{\it important fact} that the relative configuration and
relative dynamics of two HC particles is independent of
their fermionic or bosonic nature and the requirement of a
bosonic or fermionic symmetry should be enforced on the wave
functions of their ${\bf K}-$motions or spin motions.  We
use this fact in constructing $N-$particle wavefunction in
Section 2.3.  In agreement with Eqn. A-4, the SMW character
of $\zeta_k{(r)}^{\pm}$ reveals that: (i) two HC particles
have equal and opposite momenta ({\bf q},-{\bf q}) in the
frame attached to their CM which moves with momentum
${\bf K}$ in the laboratory frame, and (ii) their relative
motion maintains a center of symmetry at their CM.  This
implies
$${\bf r}_{CM}(1) = -{\bf r}_{CM}(2) =
\frac{\bf r}{2} \quad {\rm and} \quad {\bf k}_{CM}(1)
= -{\bf k}_{CM}(2) = {\bf q} \eqno(A-12)$$

\noindent
where ${\bf r}_{CM}$(i) and ${\bf k}_{CM}$(i), respectively,
refer to the position and momentum of $i-$th particle with
respect to the CM of two particles.
 
\bigskip
Since $\zeta{(r,R)}^{\pm}$ is an eigenstate of the
momentum/energy operators of the relative and CM motions of
P1 and P2 ({\it not of individual} P1/P2), it, evidently,
represents a state of their {\it mutual superposition} (MS).
Of course, one can have an alternative picture by presuming
that each of P1 and P2 after their collision at $r = 0$
falls back on the pre-collision side and assumes a kind of
{\it self superposition} (SS) state ({\it i.e,}, the
superposition of pre- and post-collision states of one
and the same particle).  Since P1 and P2 on their collision
exchange their momenta, SS state is also described by
$\zeta{(r,R)}^{\pm}$ and this agrees with the fact that
P1 and P2 are identical particles and we have no means to
ascertain whether they exchanged their positions or bounced
back after exchanging their momenta.  Evidently,
$\zeta{(r,R)}^{\pm}$ can be used to identically describe
the MS state of P1 and P2 or the SS states of P1 or P2 and
this observation helps in developing the {\it macro-orbital}
representation for a HC particle in a fluid.

\bigskip
Analyzing the 1-D motion of two HC particles, we recently,
concluded [59] that the expectation value of the relative
distance between two particles satisfies $<x> \ge
\lambda/2$.  Applying this inference to $\zeta_k(r)^{\pm}$
state of P1 and P2, we find that their $<r>$ should satisfy
$<r> \ge \lambda/2$ for ${\bf k} || {\bf r}$ case and
$<r> \ge \lambda/2\cos\theta$ for general case because the
relative dynamics of two particles, interacting through a
central force, is equivalent to such motion in 1-D.  Thus
from the experimental view point two HC particles can reach
a shortest distance, $<r>_o = \lambda/2$ and in this
situation their individual locations ({\it cf.} Eqn. A-12)
are given by $<{\bf r}_{CM}(1)>_o = - <{\bf r}_{CM}(2)>_o =
\lambda/4$.  Using similar result for their shortest possible
distance in $\phi-$space and $<V_{HC}(r)>$, {\it etc.} we
note that $\zeta_k(r)^{\pm}$ state is characterized by
$$<r> \quad \ge \quad \lambda/2 \quad  \quad {\rm or} \quad  \quad
q \quad  \ge  \quad q_o (= \pi/d),  \eqno(A-13)$$

\noindent
with $d$ being the nearest neighbor distance of two particles, 
$$<\phi> \quad \ge 2\pi,  \eqno(A-14)$$
$$<V_{HC}(r)> \quad = \quad <A\delta{(r)}>
\quad = 0, \eqno(A-15)$$

\noindent
and
$$E(2) = \quad <H(2)> \quad = \left[\frac{\hbar^2k^2}{4m} +
\frac{\hbar^2K^2}{4m}\right]. \eqno(A-16) $$
  
\noindent
Although, as evident from Eqn. A-16, two HC particles in
$\zeta{(r,R)}^{\pm}$ seems have only kinetic energy but
this does not mean that HC interaction has no impact on
$E(2)$.  We note that HC interaction controls $E(2)$
through $q \ge q_o$ (Eqn. A-13) and this has been demonstrated
in Section 2.5 and 2.6.   Analyzing the other consequences of
Eqn. A-13, we note the following :

\smallskip
\noindent
(i) {\it Quantum size :} A HC particle of momentum $q$
exclusively occupies $\lambda/2$ space if $\lambda/2 >
\sigma$ because only then the two particles maintain
$<r> \ge \lambda/2$.  We call $\lambda/2$ as
{\it quantum size} of a particle.  It is evident that
quantum size is not an absolute size of a particle, rather
it depends on the relative momentum $k = 2q$ of two
particles; as such it is the size of one particle of
momentum $q$ as seen by other particle of momentum $-q$ or
{\it vice versa}.

\smallskip
\noindent
(ii) {\it Zero-point force :} It is natural that two
particles should experience a
repulsive force if they try violate $<r> \ge \lambda/2$ and
this force would come into operation only when our system
has fixed $<r> = d$ and $\lambda/2$ is made to increase by
some suitable process.  In a many body system this happens
when the system is cooled through a temperature below which
$\lambda/2$ tend to cross $d$ or its $q$ tends to fall below
its zero-point value $q_o = \pi/d$.  We call this force a
zero-point repulsion and it can be derived from zero-point
energy of the particle $\varepsilon_o = h^2/8md^2$
corresponding $q=q_o$.  Since this force operates only when
the two particles tend have a distance shorter than
$\lambda/2$ and it does not depend of $\sigma$, it appears
that this force represents nothing but the HC repulsion
whose operational range gets extended from $\sigma$ to
$\lambda/2$ due to wave nature of particles.

\bigskip
(iii) {\it Macro-orbital representation :} Since two HC
particles in $\zeta{(r,R)}^{\pm}$ state either experience
mutual repulsion (if they somehow have $<r> < \lambda/2$),
or no force (if $<r> \ge \lambda/2$), it is clear that
each of them retains its independent identity.  Evidently,
each particle can be identified to be in its SS state
which is represented by a kind of pair waveform $\xi \equiv
\zeta^{\pm}(r,R)$ proposed to be known as
{\it macro-orbital} described by
$$\xi_i = \sqrt{2}\sin[({\bf q}_i.{\bf r}_i)]
\exp({\bf K}_i.{\bf R}_i),
\eqno(A-17)$$

\noindent
with $i$ ($i$ = 1 or 2) referring to one of the two particles
and $r_i$ being equivalent to $r_{CM}(i)$ (Eqn. A-12) varying
from $r_i = 0$ to $r_i = \lambda/2$, while $R_i$ being the CM
position of $i-$th particle.  It is evident that each
particle in its {\it macro-orbital} representation has two
motions, $q$ and $K$.  The plane wave $K-$motion can have any
$K$ ranging from 0 to $\infty$, while the $q$ is constrained
to satisfy Eqn. A-13 due to HC interaction.  A {\it
macro-orbital} identifies each particle as an entity of size
$\lambda/2$ moving with momentum $K$.  Since
$\zeta{(r,R)}^{\pm}$ is an eigenfunction of the
energy/momentum operator of the pair {\it not of individual
particle}, each particle shares the pair energy $E(2)$
(Eqn. A-16) equally and we have
$$E_1 = E_2 = \frac{E(2)}{2} = \frac{\hbar^2q^2}{2m} +
\frac{\hbar^2K^2}{8m}  \eqno(A-18) $$

\noindent
This evidently shows that each particle for its $K-$motion
behaves as an entity of mass $4m$.

\newpage

\bigskip
\noindent 
{\bf Figure -1} :  $t = T/T_{\lambda}$ dependence of
different representatives of the order parameter of
$\lambda-$transition, {\it viz.}, $n_{K=0}(T^*)$
(Eqn. 37: {\it Curve}-A$^*$), $n_{K=0}(T) = [1 -
(T/T_{\lambda})^{3/2}]$ ({\it Curve}-A), $n^*(T)$ =
[1 -$N^*(T)/N^*(T_{\lambda}$] (Eqn. 58: {\it Curve}-B),
order parameter of second order phase transition
$\Omega{(T^*)}$ (Eqn. 36: {\it Curve}-C$^*$),
$\Omega{(T)}$ (Eqn. 35: {\it Curve}-C), experimental
$He-He$ bond strain ($d_T - d_{\lambda})/(d_o -
d_{\lambda})$ ({\it Curve}-E1) and experimental
$\rho_s(t)/\rho$ ({\it Curve}-E2).

\bigskip
\begin{figure}
\centering
\includegraphics[angle = -90, width = 1.0\textwidth]{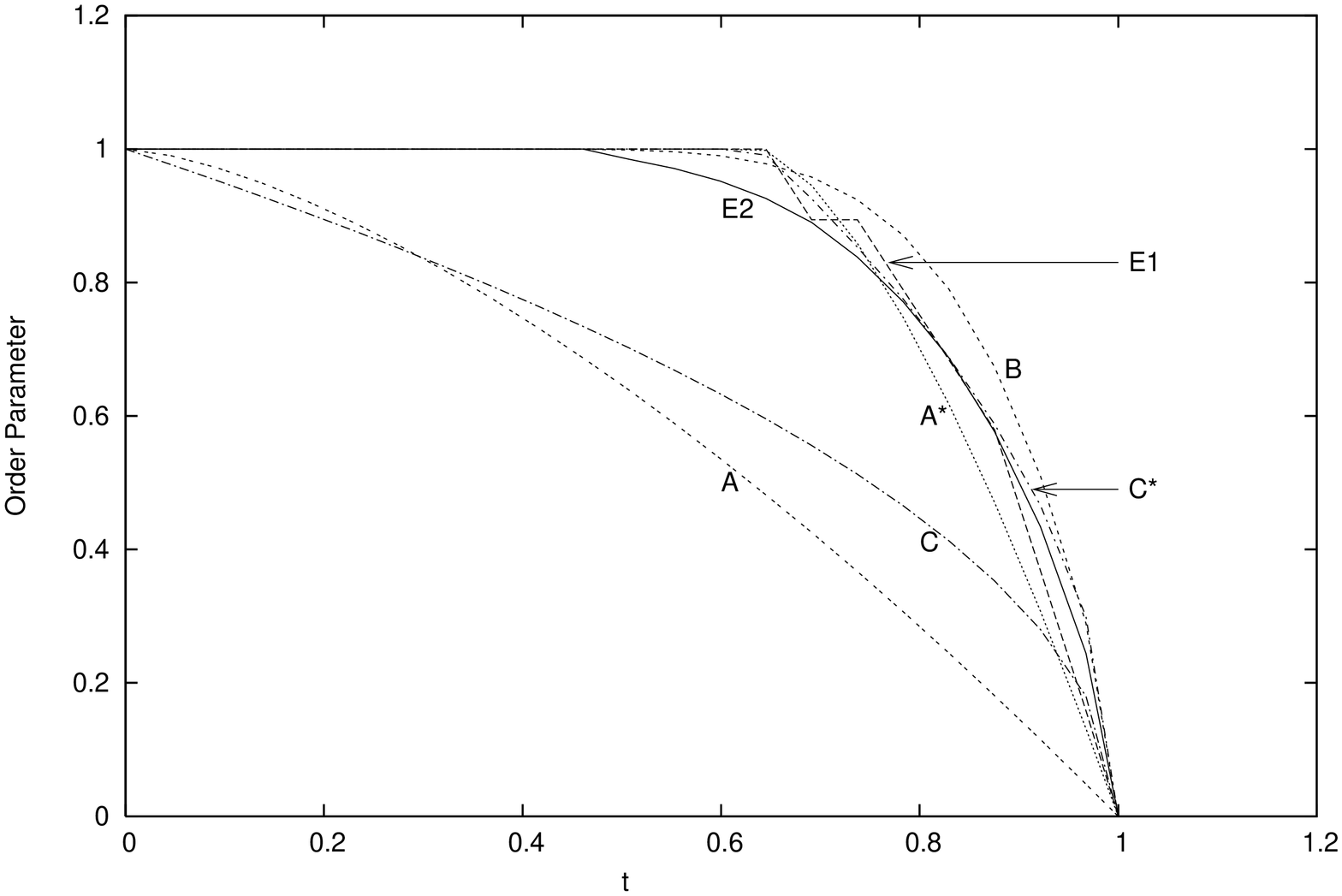}
\end{figure}


\begin{thebibliography}{??}
\bibitem[76]{Kn:gnus}    
This is evident from the fact that collective effect of binding energy 
per particle, $-E_g(T)/N$ (being $<< k_BT_o$) due to phase correlation 
among particles of $\xi{(T)}^3$ volume ($\xi{(T)}$ = coherence length,
Eqn.62) turns out be 
 $E_b = -(\xi{(T)}^3/d^3).(E_g(T)/N) 
 \approx -4{\pi}.k_BT_o.(\xi/d)$, 
 showing $|E_b| >> k_B.T_{\lambda}$; note that $\xi$ for LHE-4
 is found to vary from $\approx 10^2.d$ at $T=0$ to $\infty$ at 
 $T_{\lambda}$.
\end{thebibliography}
\end{document}